 \documentclass[preprint,review,12pt]{elsarticle}
\makeatletter
\def\ps@pprintTitle{%
 \let\@oddhead\@empty
 \let\@evenhead\@empty
 \def\@oddfoot{}%
 \let\@evenfoot\@oddfoot}
\makeatother

 \usepackage{graphicx}
\usepackage{epsfig}
\usepackage{xcolor}
\usepackage{amssymb}
\usepackage{amsmath}
\usepackage{multirow}
\usepackage{url}

\biboptions{sort&compress}
\long\def\comment#1{} 
\usepackage{xspace}
\usepackage{textcomp}
\usepackage{gensymb}
\newcommand{\ie}[0]{{\em i.e.}\ }

\newcommand{\etc}[0]{{\em etc}\xspace}

\newcommand{\vv}[0]{{\em vice-versa}\xspace}

\begin{document}
\begin{frontmatter}
\title{A proto-object based audiovisual saliency map} 
\author{Sudarshan Ramenahalli}
\address{Department of Electrical and Computer Engineering, Johns Hopkins University, Baltimore, MD \\ sramena1@jhu.edu, sudarshan.rg@gmail.com}



\begin{abstract}
  Natural environment and our interaction with it is essentially multisensory, 
where we may deploy visual, tactile and/or auditory senses to perceive, learn 
and interact with our environment. Our objective in this study is to develop a 
scene analysis algorithm using multisensory information, specifically vision and 
audio. We develop a proto-object based audiovisual saliency map (AVSM) for the 
analysis of dynamic natural scenes. A specialized audiovisual camera with $360 
\degree$ Field of View, capable of locating sound direction, is used to collect 
spatiotemporally aligned audiovisual data. We demonstrate that the performance 
of proto-object based audiovisual saliency map in detecting and localizing 
salient objects/events is in agreement with human judgment. In addition, the 
proto-object based AVSM that we compute as a linear combination of visual and 
auditory feature conspicuity maps captures a higher number of valid salient 
events compared to unisensory saliency maps. Such an algorithm can be useful in 
surveillance, robotic navigation, video compression and related applications.

\end{abstract}

\end{frontmatter}

\section{Introduction}
\label{proto-avsm-intro}

Scientists and engineers have traditionally separated the analysis of a 
multisensory scene into its constituent sensory domains. In this approach, for 
example, all auditory events are processed separately and independently of 
visual and/or somatosensory streams even though the same multisensory event 
might have created those constituent streams. It was previously necessary to 
compartmentalize the analysis because of the sheer enormity of information as 
well as the limitations of experimental techniques and computational resources. 
With recent advances, it is now possible to perform integrated analysis of 
sensory systems including interactions within and across sensory modalities. 
Such efforts are becoming increasingly common in cellular neurophysiology, 
imaging and psychophysics studies 
\cite{stein2008multisensory,stevenson2009audiovisual,
calvert2004handbook,spence2009crossmodal,alais2004ventriloquist}. 

Recent evidence from 
neuroscience~\cite{ghazanfar2006neocortex,stein2008multisensory} suggests that 
the 
traditional view that the low level areas of cortex are strictly unisensory, 
processing sensory information independently, which is later on merged in higher 
level associative areas is increasingly becoming obsolete. This has been proved 
by many fMRI~\cite{von2005interaction,watkins2006sound}, 
EEG~\cite{van2005visual} and neuro-physiological 
experiments~\cite{ghazanfar2005multisensory,wang2008visuo} at various neural 
population scales. There is now enough evidence to suggest an interplay of 
connections between thalamus, primary sensory and higher level association areas 
which are responsible for audiovisual integration. The broader implications of 
these biological findings may be that learning, memory and intelligence are 
tightly associated with the multi-sensory nature of the world.

Hence, incorporating this knowledge in computational algorithms can lead to 
better scene understanding and object recognition for which there is a great 
need. Moreover, combining visual and auditory information to associate visual 
objects with their sounds can lead to better understanding of events. For 
example, discerning whether the bat hit the baseball during a swing of the bat, 
tracking objects under severe occlusions, poor lighting conditions \etc can be 
more accurately performed only when we take audio and visual counterparts 
together. The applications of such technologies are numerous and in varied 
fields. 

In summary, a better understanding of interaction, information integration, and 
complementarity of information across senses may help us build many intelligent 
algorithms for scene analysis, object detection and recognition, human activity 
and gait detection, elder/child care and monitoring, surveillance, robotic 
navigation, biometrics {\em etc,} with better performance,  stability and 
robustness to noise. In one application, for example, fusing auditory (voice) 
and visual (face) features improved the performance of speaker identification 
and face recognition systems \cite{ccetingul2006multimodal,tamura2005toward}. 
Hence, our objective in this study is to develop a scene analysis algorithm 
using multisensory information, specifically vision and audio. We develop a 
purely bottom-up, proto-object based audiovisual saliency map (AVSM) for the 
analysis of dynamic natural scenes. 

Building on the work of \citet{russell2014model}, we add visual motion 
(Section~\ref{subsubsec:proto-avsm-feat-comp-vis-M}) as another independent 
feature type along with color, intensity and orientation, all of which undergo a 
grouping process (Section~\ref{subsec:proto-avsm-multi-res-G-pyr}) to form 
proto-objects of each feature type. In the auditory domain, we consider the 
location and intensity of sound as the only proto-objects as these are found to 
be most influential in drawing the spatial attention of an observer in many 
psycho-physics studies. Various methods of combination of the auditory and 
visual proto-object features are considered 
(Section~\ref{subsec:proto-avsm-conspic-map-combination}). We demonstrate the 
efficacy of the AVSM in predicting salient locations in the audiovisual 
environments by testing it on real world AV data collected from a specialized 
hardware (Section~\ref{sec:proto-avsm-datamethods}) that can collect $360^{0}$ 
audio and video that are temporally synchronized and spatially co-registered. 
The AVSM captures nearly all visual, auditory and audio-visually salient events, 
just as any human observer would notice in that environment. 

\section{Related Work}
\label{sec:proto-avsm-relwork}
The study of multi-sensory 
integration~\cite{alais2010multisensory,calvert2004handbook,ghazanfar2006neocortex}, 
specifically audio-visual 
integration~\cite{meredith1986visual,spence2009crossmodal} has been an active 
area of research in neuroscience, psychology and cognitive science. In the 
computer science and engineering fields, there is an increased interest in the 
recent times~\cite{evangelopoulos2008movie,song2013effet,ramenahalli2013audio}. 
For a detailed review of neuroscience and psychophysics research related to 
audio-visual interaction, please refer to 
\cite{alais2010multisensory,calvert2004handbook}. Here, we restrict our review 
to models of perception in audio-visual environments and some application 
oriented research using audio and video information. 

In one of the earliest works~\cite{grossberg1997neural}, a one-dimensional 
computational neural model of saccadic eye movement control by Superior 
Colliculus (SC) is investigated. The model can generate three different types of 
saccades: visual, multimodal and planned. It takes into account different 
coordinate transformations between retinotopic and head-centered coordinate 
systems, and the model is able to elicit multimodal enhancement and depression 
that is typically observed in SC neurons 
\cite{meredith1996spatial,meredith1987determinants}. However, the main focus is 
on Superior Colliculus function rather than studying audio-visual interaction 
from a salience perspective. A detailed model of the SC is presented in 
\cite{casey2012audio} with the aim of localizing audio-visual stimuli in real 
time. The model consists of 12,240 topographically organized neurons, which are 
hierarchically arranged into 9 feature maps. The receptive field of these 
neurons, which are fully connected to their input, are obtained through 
competitive learning. Intra-aural level differences are used to model auditory 
localization, while simple spatial and temporal differencing is used to model 
visual activity. A spiking neuron model~\cite{huo2009adaptation} of audio-visual 
integration in barn owl uses Spike Timing Dependent Plasticity (STDP) to 
modulate activity dependent axon development, which is responsible for aligning 
visual and auditory localization maps. A neuromorphic implementation of the same 
using digital and analog mixed Very Large Scale Integration (mixed VLSI) can be 
found in \cite{huo2012adaptive}.

In another neural model~\cite{anastasio2000using,patton2002multimodality} the 
visual and auditory neural inputs to the deep SC neuron are modeled as Poisson 
random variables. Their hypothesis is that the response of SC neurons is 
proportional to the presence of an audio-visual object/event in that spatial 
location which is conveyed to topographically arranged deep SC neurons via 
auditory and visual modalities. The model is able to elicit all properties of 
the SC neurons. An information theoretic explanation of super-additivity and 
other phenomena is given in a \cite{patton2002multimodality}. They also show 
that addition of a cue from another sensory modality increases the certainty of 
a target's location only if the input from initial modality/ies cannot reduce 
the uncertainty about target. Similar models are proposed in 
\cite{patton2003modeling} and in \cite{colonius2004aren}, where the problem is 
formulated based on Bayes likelihood ratio. An important 
work~\cite{ma2006bayesian} based on Bayesian inference explains a variety of cue 
combination phenomena including audio-visual spatial location estimation. 
According to the model, neuronal populations encode stimulus information using 
probabilistic population codes (PPCs) which represent probability distributions 
of stimulus properties of any arbitrary distribution and shape. They argue that 
neural populations approximate the Bayes rule using simple linear combination of 
neuronal population activities. 

In \cite{Wilson2002}, audiovisual arrays for untethered spoken interfaces are 
developed. The arrays localize the direction and distance of an auditory source 
from the microphone array, visually localize the auditory source, and then 
direct the microphone beamformer to track the speaker audio-visually. The method 
is robust to varying illumination and reverberation, and the authors report 
increased speech recognition accuracy using the AV array compared to non-array 
based processing.

In \cite{torres2014influence} the authors found that emotional saliency conveyed 
through audio, drags an observer's attention to the corresponding visual object, 
hence people often fail to notice any visual artifacts present in the video, 
suggest to exploit this property in intelligent video compression. For the same 
goal authors of \cite{lee2011efficient} implement an efficient video coding 
algorithm based on the audio-visual focus of attention where sound source is 
identified from the correlation between audio and visual motion information. The 
same premise that audio-visual events draw an observer's attention is the basis 
for their formulation. A similar approach is applied to High Definition video 
compression in \cite{rerabek2014audiovisual}. In these studies, spatial 
direction of sound was not considered, instead stereo or mono audio track 
accompanying the video was used in all computational and experimental work. 

In \cite{ruesch2008multimodal}, a multimodal bottom-up attentional system 
consisting of a combined audio-visual salience map and selective attention 
mechanism is implemented for the humanoid robot iCub. The visual salience map is 
computed from color, intensity, orientation and motion maps. The auditory 
salience map consists of the location of the sound source. Both are registered 
in ego-centric coordinates. The audio-visual salience map is constructed by 
performing a pointwise {\em max} operation on visual and auditory maps. In an 
extension to multi-camera setting~\cite{schauerte2009multi}, the 2D saliency 
maps are projected into a 3D space using ray tracing and combined as a fuzzy 
aggregations of salience spaces. In 
\cite{schauerte2011multimodal,schauerte2016bottom}, after computing the audio 
and visual saliency maps, each salient event/proto-object is parameterized by 
salience value, cluster center (mean location), and covariance matrix 
(uncertainty in estimating location). The maps are linearly combined based on 
\cite{2007OnatIntAVovertAtt}. Extensions of this approach can be found in 
\cite{kuhn2012modular}. A work related to~\cite{kuhn2012modular} is presented 
in~\cite{kuhn2012multimodal} where weighted linear combination of proto-object 
representations obtained using mean-shift clustering is detailed. Even though 
the method uses linear combination, the authors do not use motion information in 
computing the visual saliency map. A Self Organizing Map (SOM) based model of 
audio-visual integration was presented in \cite{bauer2012som} in which the 
transformations between sensory modalities, and the respective sensory 
reliabilities are learned in an unsupervisory manner. A system to detect and 
track a speaker using a multi-modal, audio-visual sensor set that fuses visual 
and auditory evidence about the presence of a speaker using Bayes network was 
presented in \cite{viciana2014audio}. In a series of papers~\cite{ 
evangelopoulos2008movie,evangelopoulos2008audiovisual,rapantzikos2007audio} 
audio-visual saliency is computed as a linear mixture of visual and auditory 
saliency maps for the purpose of movie summarization and key frame detection. No 
spatial information about audio is considered. The algorithm performs well in 
summarizing the videos for informativeness and enjoyability for movie clips of 
various genres. An extension of these models incorporating text Saliency can be 
found in~\cite{evangelopoulos2013multimodal}. By assuming a single moving sound 
source in the scene, audio was incorporated into the visual saliency map in 
\cite{nakajima2014incorporating} where sound location was associated with the 
visual object by correlating sound properties with the motion signal. By 
computing Bayesian surprise as in \cite{Itti_Baldi08}, the authors in 
\cite{nakajima2015visual} present a visual attention model driven by auditory 
cues, where surprising auditory events are used to select synchronized visual 
features and emphasize them in a audio-visual surprise map. A real-time 
multi-modal home entertainment system~\cite{korchagin2011just} performing a 
Just-In-Time association of features related to a person from audio and video 
are fused based on the shortest distance between each of the faces (in video) 
and the audio direction vector. In an intuitive study~\cite{hershey2000audio} 
speaker localization by measuring the audio-visual synchrony in terms of mutual 
information between auditory features and pixel intensity change is considered. 
In a single active speaker scenario, they obtain good preliminary results. No 
microphone arrays are used for the localization task. In \cite{blauth2012voice} 
visually detected face location is used to improve the speaker localization 
using a microphone array. A fast audiovisual attention model for human detection 
and localization is proposed in \cite{ratajczak2016fast}.

The effect of sound on gaze behavior in videos was studied in 
\cite{song2013effet,song2012different} where a preliminary computational model 
to predict eye movements was proposed. They use motion information to detect 
sound source. High level features such as face are hand labeled. A comparison of 
eye movements during visual only and audio-visual conditions with their model 
shows that adding sound information improved predictive power of their model. 
The role of salience, faces and sound in directing the attention of human 
observers (measured by gaze tracking) was studied with psychophysics experiments 
and computational modeling in \cite{coutrot2014saliency} and an audiovisual 
attention model for natural conversation scenes was proposed in 
\cite{coutrot2014audiovisual}, where they use a speaker diarization algorithm to 
compute saliency\footnote{ Speaker diarization deals with the segmentation of 
speech into non-overlapping homogeneous regions separated by silence and 
assigning each of the segmented speech bits to unique speakers. Even though 
multi-modal speaker diarization methods perform audio-visual integration, for 
example in \cite{noulas2012multimodal} they combine audio and visual information 
using an Expectation Maximization (EM) algorithm in a Dynamic Bayesian Network 
(DBN) framework, the application is specific to speech and cannot generalize to 
a saliency map, hence not reviewed. A review of these methods can be found in 
\cite{miro2012speaker}.}. Even though their study is restricted to conversation 
of humans and not applicable any generic audio-visual scene, hence cannot be 
regarded as a generalized model of audio-visual saliency, some interesting 
results are shown. Using EM algorithm to determine the individual contributions 
of bottom-up salience, faces and sound in gaze prediction, they show adding 
original speech to video improves gaze predictability, whereas adding irrelevant 
speech or unrelated natural sounds has no effect. By using speaker diarization 
algorithm~\cite{coutrot2014audiovisual} when the weight for active speakers was 
increased, their audio-visual attention model significantly outperformed the 
visual saliency model with equal weights for all faces. An audio-visual saliency 
map is developed in \cite{sidaty2014towards} where features such as color, 
intensity, orientation, faces, speech are linearly combined with unequal weights 
to give different types of saliency maps depending on the presence/absence of 
faces and/or speech. It is not clear as to whether location of the sound was 
used in their approach. Plus, they do not factor in motion, which is an 
important feature while designing a saliency map for moving pictures.

\section{Description of the model}
\label{sec:proto-avsm-model}
The computation of audiovisual saliency map is similar to the computation of 
proto-object based visual saliency map for static images explained in 
\cite{Russell_etal14}, except for (i) the addition of two new feature channels, 
the visual motion channel and the auditory loudness and location channel; and 
(ii) different ways of combining the conspicuity maps to get the final saliency 
maps. Hence, whereever the computation is identical to \cite{Russell_etal14}, we 
will only give a gist of that computation to avoid repetition and detailed 
explanation otherwise.  

\begin{figure}
	\centering
	\includegraphics[width=0.9\textwidth]{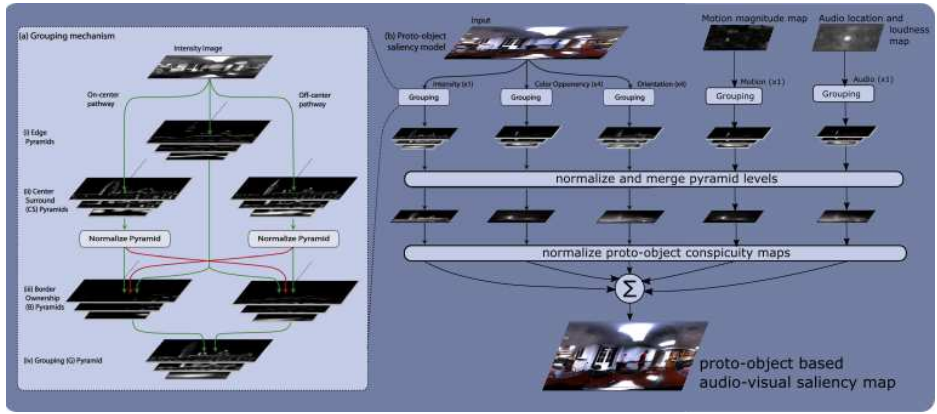}
	\caption{Computation of proto-object based AVSM. (a) Grouping mechanism: At each scale, 
feature maps are filtered with center-surround cells, normalized, followed by 
border ownership (BO) computation. Grouping cells, at each scale, receive a 
higher feed-forward input from BO cells if they are consistent with Gestalt 
properties of convexity, proximity and surroundedness. Hence, grouping gathers 
conspicuous BO activity of features at object centers. Each feature channel 
undergoes same computation at various scales (except Orientation, see 
explanation), only Intensity channel shown in Figure~\ref{fig:AVSM-Model}(a). 
(b) Audivisual saliency computation mechanism: Five feature types are 
considered: Color, Orientation, Intensity and Motion in the visual domain; 
spatial location and loudness estimate in the auditory domain. All features 
undergo grouping as explained in (a) to obtain proto-object pyramids. The 
proto-object pyramids are normalized, collapsed to get the feature specific 
conspicuity maps such that isolated strong activity is accentuated and 
distributed weak activity is suppressed. The conspicuity maps are then combined 
in different ways as explained in 
Section~\ref{subsec:proto-avsm-conspic-map-combination} to get different 
saliency maps. Figure.~5 of \cite{Russell_etal14} modified with permission.
}
	\label{fig:AVSM-Model}
\end{figure}

The AVSM is computed by grouping auditory and visual bottom-up features at 
various scales, then normalizing the grouped features within and across scales, 
followed by merging features across scales and linear combination of the 
resulting feature conspicuity maps (Figure~\ref{fig:AVSM-Model}). The features 
are derived from the color video and multi-channel audio input 
(Section~\ref{sec:proto-avsm-datamethods}) without any top-down attentional 
biases, hence the computation is purely bottom-up. And the mechanism of grouping 
``binds'' features within a channel into candidate objects or ``proto-objects''. 
Approximate size and location are the only properties of objects that the 
grouping mechanism estimates, hence they are termed ``proto-objects''. Such 
proto-objects, many of them, form simultaneously and dissolve 
rapidly~\cite{rensink2000dynamic} in a purely bottom-up manner. Top-down 
attention is required to hold them together into coherent objects. 

Also, computation of AVSM is completely feed-forward. Many spatial scales are 
used to achieve scale invariance.  First the independent feature maps are 
computed, features within each channel are grouped into proto-objects. Such 
proto-object feature pyramids at various scales are normalized within and across 
scales. Such feature pyramids are merged across scales followed by normalization 
across feature channels to give rise to conspicuity maps. The conspicuity maps 
are linearly combined to get the final AVSM.  Each of these steps is explained 
in more detail below. 

\subsection{Computation of feature channels}
\label{subsec:proto-avsm-feat-comp}
We consider color, intensity, orientation and motion as separate, independent  
feature channels in the visual domain. Loudness and spatial location as features 
in the auditory space. The audio-visual camera equipment used for data gathering 
(See Section~\ref{sec:proto-avsm-datamethods}) guarantees spatial and temporal 
concurrency of audio and video. 

A single intensity channel, where intensity is computed as the average of Red, 
Green and Blue color channels is used. Four feature sub-channels for angle, 
$\theta = \{0,\frac{\pi}{4}, \frac{\pi}{2}, \frac{3\pi}{4}\}$ are used for 
Orientation channel. Four color opponency feature sub-channels: Red-Green 
($\mathcal{RG}$), Green-Red ($\mathcal{GR}$), Blue-Yellow ($\mathcal{BY}$) and 
Yellow-Blue ($\mathcal{YB}$) are used for color channel. The computation of 
color, orientation and intensity feature channels is identical to 
\citet{Russell_etal14}.

Visual motion, auditory loudness and location estimate are the newly added 
features, the computation of which is explained below. 

\subsubsection{Visual motion channel}
\label{subsubsec:proto-avsm-feat-comp-vis-M}
Motion is computed using the optical flow algorithm described in 
\cite{sun2010secrets} and the corresponding code available at 
\cite{opticFlowDSun}. Consider two successive video frames, $I(x,y,t)$ and $I(x, 
y, t + 1)$. If the underlying object has moved between $t$ and $t + 1$, then the 
intensity at pixel location, $(x,y)$ at time, $t$ should be the same in a nearby 
pixel location at $(x + \Delta x, y + \Delta y)$ in the successive frame at $t + 
1$. Using this as one of the constraints, the flow is estimated which gives the 
horizontal and vertical velocity components, $u(x,y,t)$ and $v(x,y,t)$ 
respectively, at each pixel location $(x,y)$ at time $t$. For a more detailed 
explanation, see \cite{sun2010secrets}. Since we are interested in detecting 
salient events only, we do not take into account the exact motion at each 
location as given by $u(x,y,t)$ and $v(x,y,t)$, instead, we look at how big the 
motion is at each location in the image. The magnitude of motion at each 
location is computed as, 

\begin{equation}
M(x,y,t) = \sqrt{u(x,y,t)^{2} + v(x,y,t)^{2}}
\label{eq:proto-avsm-feat-comp-vis-MotChannel}
\end{equation}

\begin{figure}
	\centering
	\includegraphics[width=\linewidth]{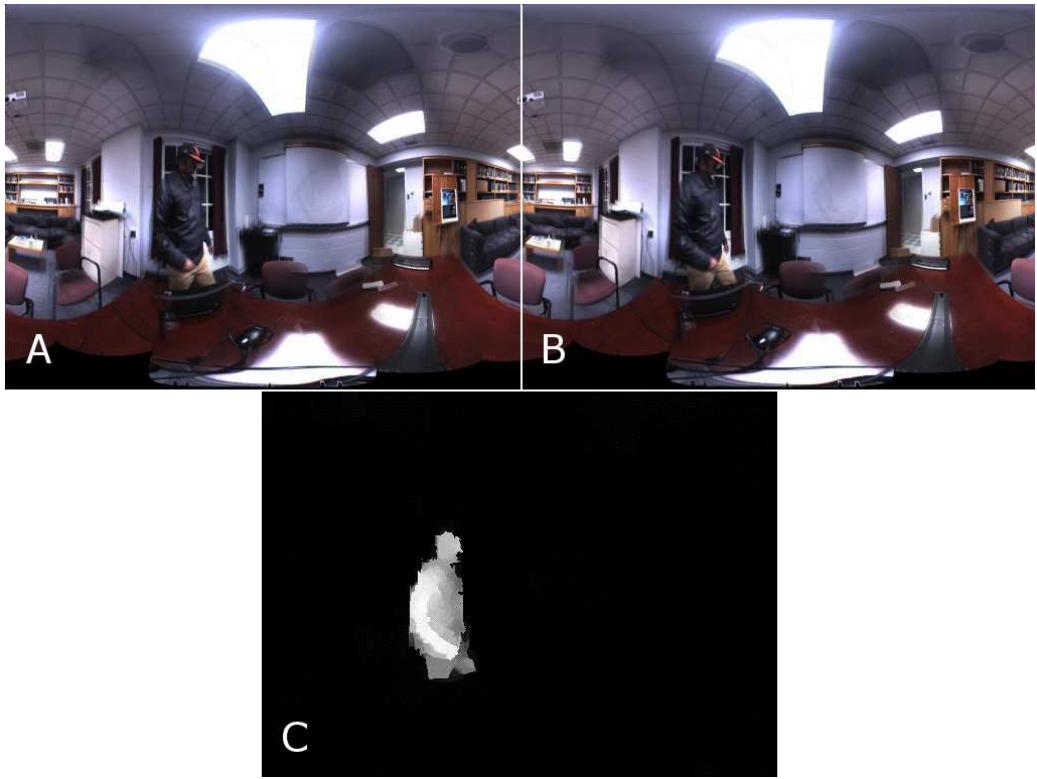}
	\caption{Computation of motion magnitude map. (A)A frame from the dataset, (B)the 
successive frame. The person in the two frames is the only moving object in the 
two image frames, moving from left to right in the image. (C)The motion 
magnitude map
}
	\label{fig:proto-avsm-feat-comp-vis-MotChannel}
\end{figure}

The motion map, $M(x,y,t)$ gives the magnitude of motion at each location in the 
visual scene at different time instances, $t$. 
Figure~\ref{fig:proto-avsm-feat-comp-vis-MotChannel} shows two successive frames 
of the video in (A) and (B) respectively. The computed magnitude of 
motion\footnote{for an alternate method to compute motion and incorporate it 
into saliency map, see \cite{molin2013proto}} using the optic flow method in 
\cite{sun2010secrets} is shown in (C). The person in the video is the only 
moving source. 

\subsubsection{Auditory loudness and location channel}
\label{subsubsec:proto-avsm-feat-comp-aud-LL}
\begin{figure}
	\centering
	\includegraphics[width=\linewidth]{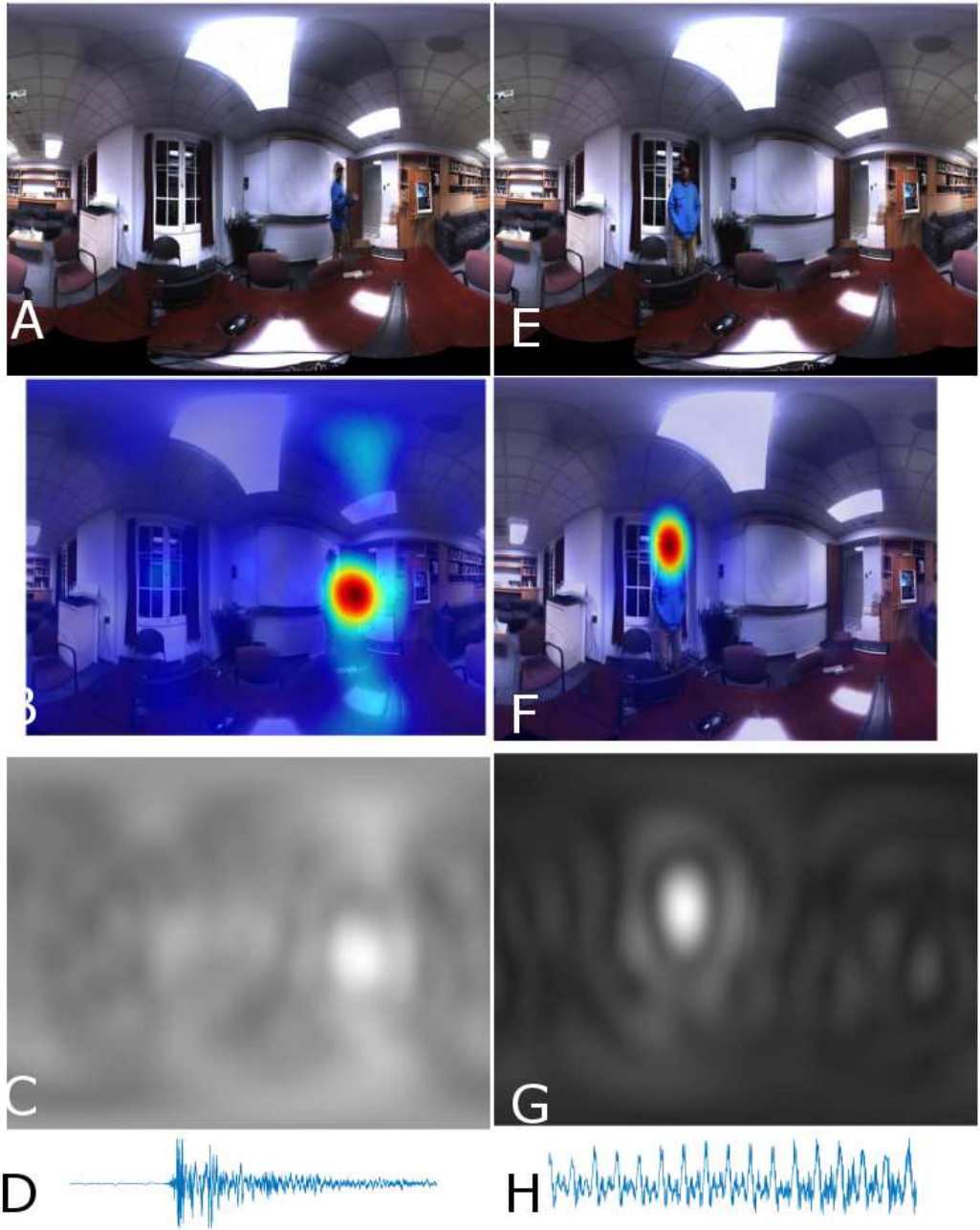}
	\caption{Computation of auditory location and loudness map: Two different video frames in 
(A) and (E). Corresponding audio samples from one of the audio channels are 
shown in (D) and (H), which display only 4410 samples, \ie, audio of 0.1 s. 
Corresponding auditory loudness and location maps are in (C) and (G). 
Corresponding video overlays in (B) and (F) for illustration only, not used in 
any computation.  Left column: sound of a clap. Right column: some part of the 
word ``eight'' uttered by the person
}
	\label{fig:proto-avsm-feat-comp-aud-LL}
\end{figure}

The auditory input consists of a recording of the 3D sound field using 64 
microphones arranged on a sphere (See Section~\ref{sec:proto-avsm-datamethods} 
for details). We compute a single map for both loudness and location of sound 
sources, $A(x,y,t)$. The value at a location in the map, $A(x,y,t)$ gives an 
estimate of loudness at that location at time, $t$, hence we simultaneously get 
the presence and loudness of sound sources at every location in the entire 
environment. These two features are computed using beamforming technique as 
described in \cite{donovan2007microphone,meyer2002highly,o2007real}. A more 
detailed account is given in Section~\ref{sec:proto-avsm-datamethods}. Two 
different frames of video and the corresponding auditory loudness and location 
maps superimposed on the visual images are shown in 
Figure~\ref{fig:proto-avsm-feat-comp-aud-LL}(B) and (F) respectively. Warm 
colors indicate higher intensity of sound from that location in the video.

\subsection{Feature pyramid decomposition}
\label{subsec:proto-avsm-multi-res-pyr}
Feature pyramids are computed for each type. As the scale increases, the 
resolution of the feature map decreases. The feature maps of successively higher 
scales are computed by downsampling the feature map from the previous scale. The 
downsampling factor can be either $\sqrt{2}$ (half-octave) or 2 (full octave). 
The feature pyramids thus obtained are used to compute proto-objects by border 
ownership and grouping computation process explained the next two sections. 

\subsection{Border ownership pyramid computation}
\label{subsec:proto-avsm-multi-res-BO-pyr}
Computation of proto-objects by grouping mechanism can be divided into two 
sub-steps: (i) border ownership pyramid computation, and (ii) grouping pyramid 
computation. 

The operations performed on any of the features, auditory or visual is the same. 
Edges of four orientations, $\theta = \{0, \pi/4, \pi/2, 3\pi/4\}$ are computed 
using the Gabor filter bank. The V1 complex cell responses~\cite{Russell_etal14} 
thus obtained are used to construct the edge pyramids. Border ownership response 
is computed by modulating the edge pyramid by the activity of center-surround 
feature differences on either side of the border. The rationale behind this is 
the observation made by \citet{Zhang_vonderHeydt10}, where they reported that 
the activity border ownership cells was enhanced when image fragments were 
placed on their preferred side, but suppressed for the non-preferred side. 

Two types of center-surround (CS) feature pyramids are used. The center-surround 
light pyramid detects strong features surrounded by weak ones. Similarly, to 
detect weak features surrounded by a strong background, a center-surround dark 
pyramid is used. The rationale behind light and dark CS pyramids is that there 
can be bright objects on a dark background and \vv. The center-surround pyramids 
are constructed by convolving feature maps with Difference of Gaussian (DoG) 
filters. 

The CS pyramid computation is performed in this manner for all feature types 
including motion and audio, except for the orientation channel. For the 
orientation feature channel, the  DoG filters are replaced by the even symmetric 
Gabor filters which detect edges. This is because, for the orientation channel, 
feature contrasts are not typically symmetric as in the case of other channels, 
but oriented at a specific angle. 
 
An important step in the border ownership computation is normalization of the 
center-surround feature pyramids. We follow the noramlization method used in 
\cite{Itti_etal98a} which enhances isolated high activities and suppresses many 
closely clustered similar activities. 

This normalization step enables comparison of light and dark CS pyramids. 
Because of the normalization, border ownership activity and grouping activity 
are proportionately modulated, deciding relative salience of proto-objects from 
the grouping activity.

The border ownership (BO) pyramids corresponding to light and dark CS pyramids 
are constructed by modulating the edge activity by the normalized CS pyramid 
activity. The light and dark BO pyramids are merged across scales and summed to 
get contrast polarity invariant BO pyramids. For each orientation, two BO 
pyramids with opposite BO preferences are be computed. From this, the winning BO 
pyramids are computed by a winner-take-all mechanism.

\subsection{Grouping pyramid computation}
\label{subsec:proto-avsm-multi-res-G-pyr}
The grouping computation shifts the BO activity from edge pixels to object 
centers. Grouping pyramids are computed by integrating the winning BO pyramid 
activity such that selectivity for Gestalt properties of convexity, proximity 
and surroundedness is enhanced. This is done by using Grouping cells in this 
computation, which have an annular receptive field. The shape of G cells gives 
rise to selectivity for convex, surrounded objects. At this stage we have the 
grouping or proto-object pyramids which are normalized and combined across 
scales to compute feature conspicuity maps, and then the saliency map. 

\subsection{Normalization and across-scale combination of grouping pyramids}
\label{subsec:proto-avsm-across-scal-G-pyr-combination}
The computation of grouping pyramids as explained in 
Section~\ref{subsec:proto-avsm-multi-res-G-pyr} is performed for each feature 
type. Let us represent the grouping pyramid for intensity feature channel by 
$\mathcal{G}^{k}_{\mathcal{I}}(x,y,t)$, where $k$ denotes the scale of the 
proto-object map in the grouping pyramid. The color feature sub-channel grouping 
pyramids are represented as $\mathcal{G}^{k}_{\mathcal{RG}}(x,y,t)$ for 
Red-Green, $\mathcal{G}^{k}_{\mathcal{GR}}(x,y,t)$ for Green-Red, 
$\mathcal{G}^{k}_{\mathcal{BY}}(x,y,t)$ for Blue-Yellow and 
$\mathcal{G}^{k}_{\mathcal{YB}}(x,y,t)$ for Yellow-Blue color opponencies. The 
orientation grouping pyramids are denoted by 
$\mathcal{G}^{k}_{\mathcal{O}}(x,y,t,\theta)$ where $\theta$ denotes 
orientation, motion feature channel by $\mathcal{G}^{k}_{\mathcal{M}}(x,y,t)$ 
and auditory location and intensity feature channel by 
$\mathcal{G}^{k}_{\mathcal{A}}(x,y,t)$. The corresponding conspicuity maps for , 
intensity $\mathcal{I}(x,y,t)$, color $\mathcal{C}(x,y,t)$, orientation 
$\mathcal{O}(x,y,t)$, motion, $\mathcal{M}(x,y,t)$ and auditory location and 
loudness estimate,  $\mathcal{A}(x,y,t)$ are respectively obtained as,

\begin{equation}
\mathcal{I}(x,y,t) = \bigoplus_{k = 1}^{k = 10} \mathcal{N}(\mathcal{G}_{\mathcal{I}}^{k}(x,y,t)) 
\label{eq:proto-avsm-Int-Cons-map}
\end{equation}

\begin{equation}
\begin{split}
\mathcal{C}(x,y,t) = \bigoplus_{k = 1}^{k = 10} \Big ( \mathcal{N}(\mathcal{G}_{\mathcal{RG}}^{k}(x,y,t)) + \mathcal{N}(\mathcal{G}_{\mathcal{GR}}^{k}(x,y,t)) \\ 
+ \mathcal{N}(\mathcal{G}_{\mathcal{BY}}^{k}(x,y,t)) + \mathcal{N}(\mathcal{G}_{\mathcal{YB}}^{k}(x,y,t)) \Big ) 
\end{split}
\label{eq:proto-avsm-Color-Cons-map}
\end{equation}

\begin{equation}
\mathcal{O}(x,y,t) = \bigoplus_{k = 1}^{k = 10} \quad \sum_{\theta \in \{ 0,\frac{\pi}{4},\frac{\pi}{2}, 3 \frac{\pi}{4} \}}  \mathcal{N}(\mathcal{G}^{k}_{\mathcal{O}}(x,y,t,\theta))
\label{eq:proto-avsm-Ori-Cons-map}
\end{equation}

\begin{equation}
\mathcal{M}(x,y,t) = \displaystyle \bigoplus_{k = 1}^{k = 10} \mathcal{N}(\mathcal{G}_{\mathcal{M}}^{k}(x,y,t))
\label{eq:proto-avsm-Mot-Cons-map}
\end{equation}

\begin{equation}
\mathcal{A}(x,y,t) = \displaystyle \bigoplus_{k = 1}^{k = 10} \mathcal{N}(\mathcal{G}_{\mathcal{A}}^{k}(x,y,t))
\label{eq:proto-avsm-Aud-Cons-map}
\end{equation}
where $\mathcal{N}(.)$ is a normalization step as explained in 
\citet{Itti_etal98a}, which accentuates strong isolated activity and suppresses 
many weak activities, the symbol $\bigoplus$ denotes ``across-scale'' addition 
of the proto-object maps, which is done by resampling (up- or down-sampling 
depending on the scale, $k$) maps at each level to a common scale (in this case, 
the common scale is $k = 8$) and then doing pixel-by-pixel addition. We use the 
same set of parameters as in Table 1 of \citet{Russell_etal14}  for our 
computation as well. 

The conspicuity maps, due to varied number of feature sub-channels have 
different ranges of activity, hence if we linearly combine without any rescaling 
to a common scale, those features with higher number of sub-channels may 
dominate. Hence, each  feature conspicuity map is rescaled to the same range, 
$[0, \ldots, 1]$. The conspicuity maps are combined in different ways to get 
different types of saliency maps as explained in 
Section~\ref{subsec:proto-avsm-conspic-map-combination}. 

\subsection{Combination of conspicuity maps}
\label{subsec:proto-avsm-conspic-map-combination}

The visual saliency map is computed as,

\begin{equation}
\begin{split}
\mathcal{VSM}(x,y,t) =   w_{I}\mathcal{R}(\mathcal{I}(x,y,t)) + w_{C} \mathcal{R}(\mathcal{C}(x,y,t)) \\ +  w_{O} \mathcal{R}(\mathcal{O}(x,y,t)) + w_{M} \mathcal{R}(\mathcal{M}(x,y,t))
\label{eq:proto-avsm-VSM-map}
\end{split}
\end{equation}
where $\mathcal{VSM}(x,y,t)$ is the visual saliency map, $\mathcal{R}(.)$ is the 
rescaling operator that rescales each map to the same range, $[0, \ldots, 1]$ 
and $w_{I}, w_{C}, w_{O}$ and $w_{M}$ are the individual weights for intensity, 
color, orientation and motion conspicuity maps, respectively. In our 
implementation, all weights are equal and each is set to 0.25, \ie, $w_{I} =  
w_{C} =  w_{O} = w_{M} = \frac{1}{4}$.

Since audio is a single feature channel, the conspicuity map for auditory 
location and loudness is also the auditory saliency map, $\mathcal{ASM}(x,y,t)$. 

We compute the audio-visual saliency map in three different ways to compare the 
most effective method to identify salient events (See 
Section~\ref{sec:proto-avsm-results-discussion} for related discussion).

In the first method a weighted combination of all feature maps is done to get 
the audio-visual saliency map as,
\begin{equation}
\begin{split}
\mathcal{AVSM}_{1}(x,y,t) =   w_{I}\mathcal{R}(\mathcal{I}(x,y,t)) + w_{C} \mathcal{R}(\mathcal{C}(x,y,t)) + w_{O} \mathcal{R}(\mathcal{O}(x,y,t)) \\ + w_{M} \mathcal{R}(\mathcal{M}(x,y,t)) + w_{A} \mathcal{R}(\mathcal{A}(x,y,t))
\label{eq:proto-avsm-AVSM1-map}
\end{split}
\end{equation}
where different weights can be set for $w_{(.)}$ such that the sum of all 
weights equals 1. In our implementation, all weights are set equal, \ie, $w_{I} 
=  w_{C} =  w_{O} = w_{M} = w_{A} = \frac{1}{5}$. 

In the second method, the visual saliency map is computed as in 
Equation~\ref{eq:proto-avsm-VSM-map} and then a simple average of the visual 
saliency map and the auditory conspicuity map (also auditory saliency map, 
$\mathcal{ASM}(x,y,t)$) is computed to get the audio-visual saliency map as,
\begin{equation}
\mathcal{AVSM}_{2}(x,y,t) =   \frac{1}{2}\Big (\mathcal{R}(\mathcal{VSM}(x,y,t)) +  \mathcal{R}(\mathcal{A}(x,y,t)) \Big )
\label{eq:proto-avsm-AVSM2-map}
\end{equation}

The distribution of weights in Equation~\ref{eq:proto-avsm-AVSM2-map} is 
different from that in Equation~\ref{eq:proto-avsm-AVSM1-map}. In method 2, a 
``late combination'' of the visual and auditory saliency maps is performed, 
which results in an increase in the weight of the auditory saliency map and a 
reduction in weights for the individual feature conspicuity maps of the visual 
domain. 

In the last method, in addition to a linear combination of the visual and 
auditory saliency maps, a product term is added as,

\begin{equation}
\begin{split}
\mathcal{AVSM}_{3}(x,y,t) =   \Big (\mathcal{R}(\mathcal{VSM}(x,y,t)) +  \mathcal{R}(\mathcal{A}(x,y,t)) \\ + \mathcal{R}(\mathcal{VSM}(x,y,t)) \otimes \mathcal{R}(\mathcal{A}(x,y,t)) \Big )
\label{eq:proto-avsm-AVSM3-map}
\end{split}
\end{equation}
where the symbol, $\otimes$ denotes a point-by-point multiplication of pixel 
values of the corresponding saliency maps. The effect of the product term is to 
increase the saliency of those events that are salient in both visual and 
auditory domains, thereby to enhance the saliency of spatiotemporally concurrent 
audiovisual events. A comparison of the different saliency maps in detecting 
salient events is in Section~\ref{sec:proto-avsm-results-discussion}. 

\section{Data and Methods}
\label{sec:proto-avsm-datamethods}
\begin{figure}
	\centering
	\includegraphics[width=\linewidth]{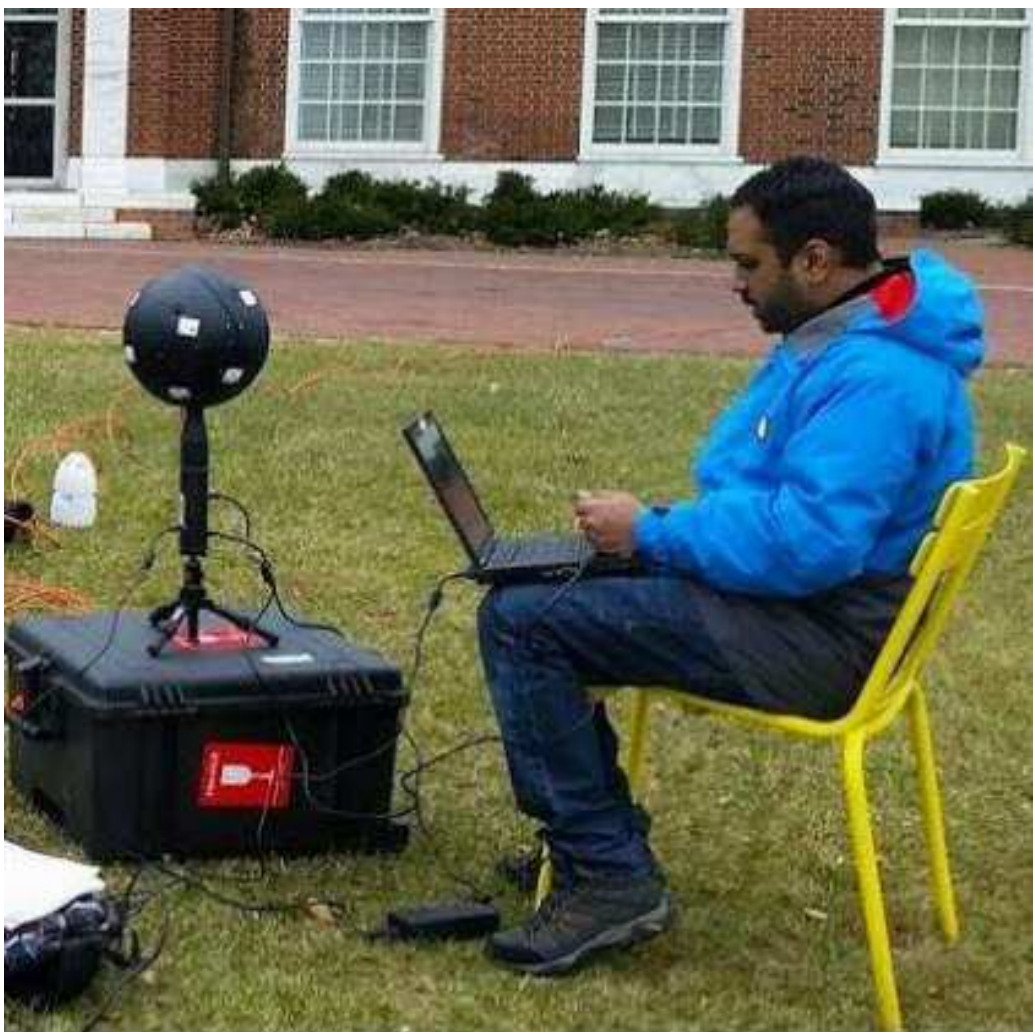}
	\caption{The Audio-Visual Camera being used to collect data in a recording session 
outdoors
}
	\label{fig:proto-avsm-AVCam-HWsetup}
\end{figure}

Audio-Visual data is collected using the VisiSonics 
RealSpace\textsuperscript{TM} audio-visual 
camera~\cite{donovan2007microphone,o2007real}. The AV camera consists of a 
spherical microphone array with 64 microphones arranged on a sphere of 8 inches 
diameter and 15 HD cameras arranged on the same sphere 
(Figure~\ref{fig:proto-avsm-AVCam-HWsetup}). Each video camera can record color 
(RGB) videos at a resolution of $1328 \times 1044$ pixels per frame and 10 
frames per second. The audio channels record high fidelity audio at a sampling 
rate of 44.1 kHz per channel. The audio and video data are converted into a 
single USB 3.0 compliant stream which is accepted by a laptop computer with 
Graphical Processing Units. The individual videos are stitched together to 
produce a panoramic view of the scene in two different projection types: 
spherical and Mercator. The audio and video streams are synchronized by an 
internal Field Programmable Gate Array (FPGA) based processor. The equipment can 
be used to localize sounds and display them on the panoramic video in real time 
and also record AV data for later analysis. The length of each recording can be 
set at any value between 10 seconds and 390 seconds . The gain for each 
recording session can set to three predefined values: -20 dB, 0 dB and +20 dB. 
This is particularly helpful to record sounds with high fidelity in indoor, 
outdoor and noisy conditions. 

To compute the loudness and location estimate of sound sources in the scene, 
Spherical Harmonics Beamforming (SHB) technique is used. The 3D sound field 
sampled at discrete locations on a solid sphere is decomposed into spherical 
harmonics, whose angular part resolves the direction of the sound field. 
Spherical harmonics are the 2D counterpart of Fourier transforms (defined on a 
unit circle) defined on the surface of a sphere. A description of the 
mathematical details of the SHB method is beyond the scope of this  article, 
interested readers can find details in 
\cite{donovan2007microphone,o2007real,meyer2002highly}. 

The 64 audio channels recorded at 44100 Hz are divided into frames, each of 4410 
samples. This gives us 10 audio frames per second, which is equal to the video 
frame rate. Spherical harmonic beamforming is done for the audio frames to 
locate sounds in the frequency range, $[300, 6500]$ Hz. For natural sounds, a 
wide frequency range, as chosen is sufficient to estimate location and loudness 
of most types of sounds including speech and music. The azimuth angle for SHB is 
chosen in the range, $[0, 2 \pi]$ with the angular resolution of $\frac{2 
\pi}{128}$ radians, \ie, $2.8125^{0}$ and the elevation angle in $0, \pi$ with 
an angular resolution of $\frac{\pi}{64}$ radians, \ie, $2.8125^{0}$ radians. 
The output of SHB is the auditory location and loudness estimate map as shown in 
Figure~\ref{fig:proto-avsm-feat-comp-aud-LL}. 

\begin{figure}
	\centering
	\includegraphics[width=\linewidth]{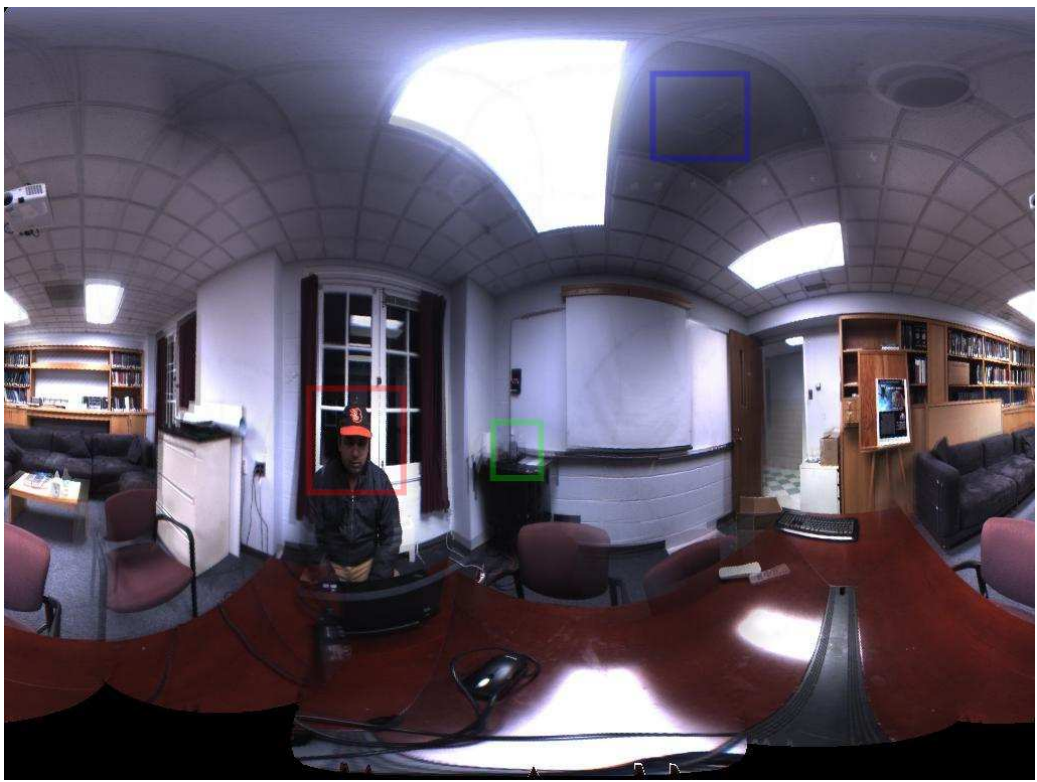}
	\caption{The audiovisual data collection scene. The scene consists of a loudspeaker 
(green box), a person and an air conditioning vent (blue box)
}
	\label{fig:proto-avsm-AVData-Setup}
\end{figure}

We collected four audiovisual datasets using the AV camera equipment, where 
three datasets are 60 seconds in length and the other one is of 120 seconds 
duration, all indoors. The AV camera equipment and our algorithms can handle 
data from any type of audiovisual surroundings, indoor or outdoor. We made sure 
all combination of salient events in purely visual intensity, color, motion, 
audio and audiovisual domains were present. SHB was performed with parameters 
set as explained in the previous paragraph to get the sound location and 
loudness estimates. The videos are stitched together to produce panoramic image 
in the Mercator projection which is used in our saliency computation. The stitch 
depth for panoramic images was 14 feet, hence in objects that are too close to 
the AV camera appear to be blurred due to overlapping of images from different 
cameras on the sphere (Figure~\ref{fig:proto-avsm-AVData-Setup}). 

The scene consisted of a loudspeaker placed on a desk in one corner of the room 
(green box) playing documentaries, a person (author) either sitting or moving 
around the AV camera equipment clapping or uttering numbers out loud, an air 
conditioning vent (blue box) making some audible noise and other objects visible 
in the scene. The loudspeaker acts like a stationary sound source, which was 
switched on/off during the recording session to produce abrupt onsets/offsets. 
The person wearing a black/blue jacket moving around the recording equipment 
acts like a salient visual object with or without motion, when speaking acts as 
an audiovisual source. The person moves in and out of the room producing abrupt 
motion onsets/offsets. The air conditioning vent, which happens to be very close 
to the recording equipment is a source of noise which is audible to anyone 
present in the room, acts as another stationary audio source. The bright lights 
present in the room act as visually salient stationary objects. The set of 
sources together produce all possible combinations of visually salient events 
with or without motion, acoustically salient sources with or without motion and 
audiovisual salient events/objects with and without motion. The dataset can be 
viewed/listened to at the following url: 
\url{https://preview.tinyurl.com/ybg4fch4}. Results of audiovisual saliency, 
comparison with unisensory saliencies on this dataset are discussed in the next 
section. 

\section{Results and Discussion}
\label{sec:proto-avsm-results-discussion}
First, we will examine which of the three audiovisual saliency computation 
methods described in Section~\ref{subsec:proto-avsm-conspic-map-combination}, 
Eqs~\ref{eq:proto-avsm-AVSM1-map} - \ref{eq:proto-avsm-AVSM3-map} performs well 
for different stimulus conditions. Then we will compare results from the best 
AVSM with the unisensory saliency maps followed by discussion of the results.

All saliency maps computed as explained in 
Section~\ref{subsec:proto-avsm-conspic-map-combination} will have salience value 
in $(0,1)$ range. On such a saliency map, unisensory or audiovisual, anything 
above a threshold of 0.75 is determined as highly salient. This threshold is 
same for all saliency maps, Visual Saliency Map ($\mathcal{VSM}$, variables 
$(x,y,t)$ dropped as unnecessary here), Auditory Saliency Map ($\mathcal{ASM}$) 
and the three different Audio-Visual Saliency Maps ($\mathcal{AVSM}_{i}, 
\text{where }i=1, 2, 3$). Hence, this provides a common baseline to compare the 
workings of unisensory SMs with AVSM, and among different AVSMs. 

\begin{figure}
	\centering
	\includegraphics[width=\linewidth]{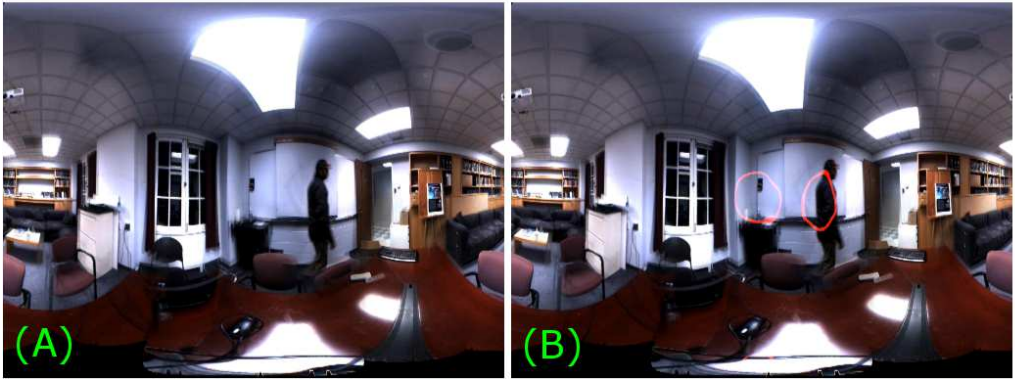}
	\caption{Visualization of results with isocontours. (A) Input frame \# 77 of Dataset 2. 
(B) The red contour superimposed on the same input frame is the isocontour of 
salience values. All the values along the red line are equal to the threshold 
value. Anything inside the closed red isocontour has a salience value greater 
than 0.75
}
	\label{fig:AVSM-Isocontour-Illus}
\end{figure}

To visualize the results we did the following: On the saliency map (can be 
$\mathcal{VSM}$, $\mathcal{ASM}$ or $\mathcal{AVSM}_{i}$), saliency value based 
isocontours for the threshold of 0.75 are drawn and superimposed on each of the 
input video frame. For example, see Figure~\ref{fig:AVSM-Isocontour-Illus}, 
where  $\mathcal{AVSM}_{1}$ for frame \# 77 of Dataset 2 is shown. Any thing 
that is inside the closed red contour of 
Figure~\ref{fig:AVSM-Isocontour-Illus}(B) is highly salient and has a saliency 
value greater than 0.75. Outside the isocontour, the salience value is less than 
0.75. Exactly, along the isocontour the salience value is 0.75 (precisely, $0.75 
\pm 0.02$). 

The results can be best interpreted by watching the input and different saliency 
map videos. But, since it is not possible to show all the frames and for the 
lack of a better way of presenting the results, we display the saliency maps for 
a few key frames only. The videos and individual frames of the saliency maps are 
available at the url: \url{https://preview.tinyurl.com/ybg4fch4}. 

\begin{figure}
	\centering
	\includegraphics[width=\linewidth]{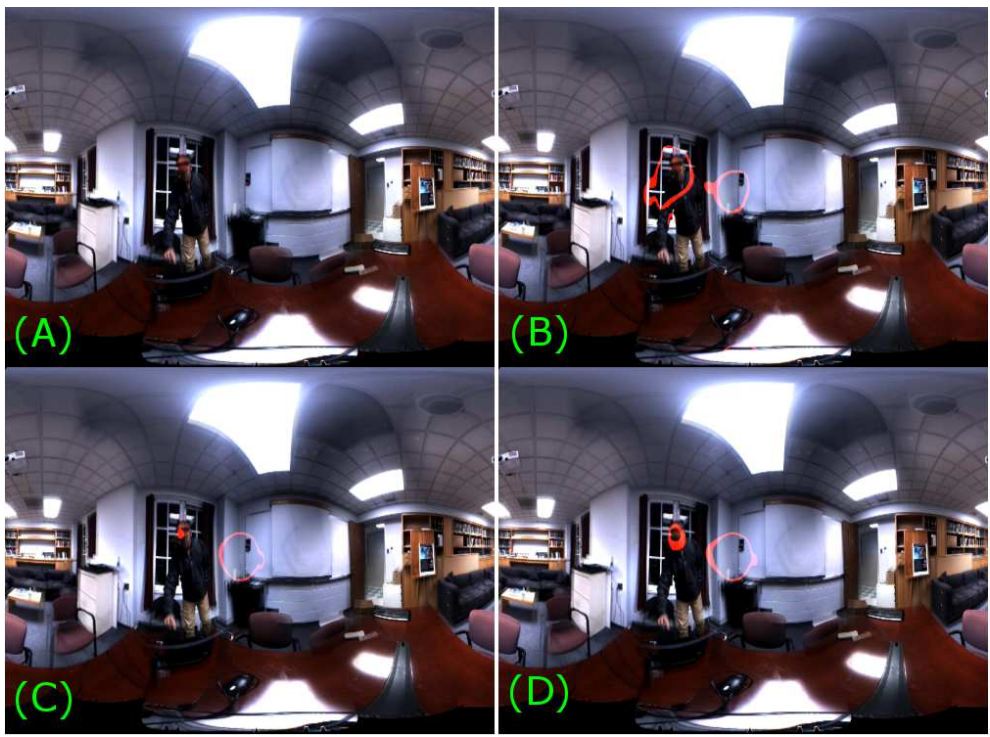}
	\caption{Comparison of audiovisual saliency computation methods. (A) Input frame \# 393 
of Dataset 1. A small loudspeaker at the corner of a room, which is located 
almost at the center of the image frame, is playing a documentary, hence 
constitutes a salient stationary auditory event. The person is leaning forward 
gives rise to salient visual motion. (B) $\mathcal{AVSM}_{1}$ computed using 
Eq~\ref{eq:proto-avsm-AVSM1-map} where all 5 feature channels are combined 
linearly with equal weights. (C) $\mathcal{AVSM}_{2}$ computed using 
Eq~\ref{eq:proto-avsm-AVSM2-map} where $\mathcal{VSM}$ and $\mathcal{ASM}$ are 
averaged. (D) $\mathcal{AVSM}_{3}$ computed using 
Eq~\ref{eq:proto-avsm-AVSM3-map}. All methods give similar results with minor 
differences
}
	\label{fig:AVSM-Comparison-1}
\end{figure}

Figure~\ref{fig:AVSM-Comparison-1} shows $\mathcal{AVSM}_{1}$, 
$\mathcal{AVSM}_{2}$ and $\mathcal{AVSM}_{3}$ for input image frame \# 393 of 
Dataset 1. At that moment in the scene, the loudspeaker (at the center of the 
image frame) was playing a documentary and the  person was moving forward. So, 
there is a salient stationary auditory event and a salient visual motion. From 
visual inspection of Figure~\ref{fig:AVSM-Comparison-1}, it is clear that 
$\mathcal{AVSM}_{1}$, $\mathcal{AVSM}_{2}$ and $\mathcal{AVSM}_{3}$  give 
roughly the same results, and are able to detect salient events in both 
modalities. This is the behavior we see in all AVSMs ($\mathcal{AVSM}_{i}$) for 
a majority of frames. But, in some cases, when the scene reduces to a static 
image, the behavior exhibited by each of the methods will be somewhat different. 

\begin{figure}
	\centering
	\includegraphics[width=\linewidth]{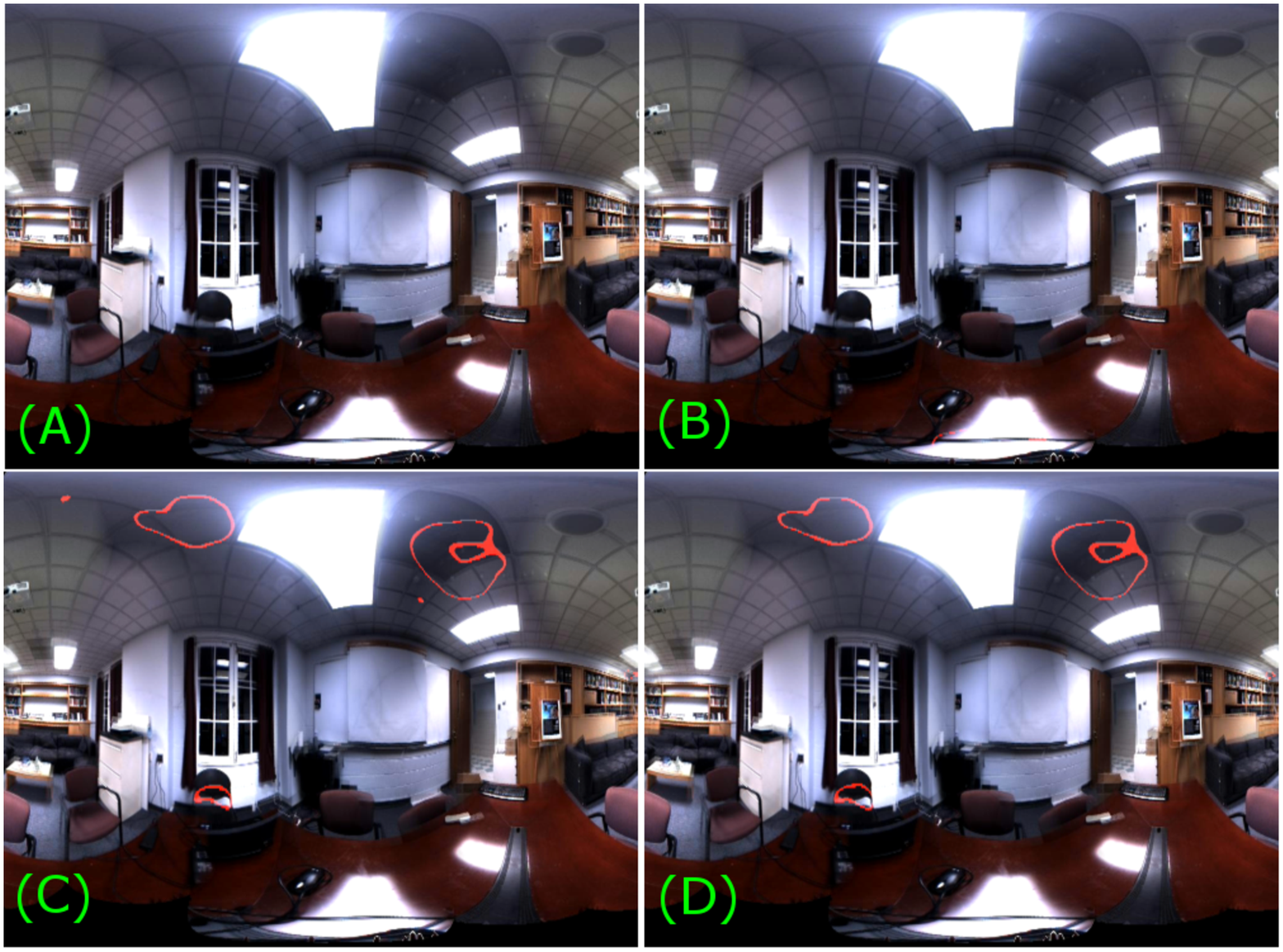}
	\caption{Audiovisual saliency in a static scene. (A) Input frame \# 173 of Dataset 3. The 
scene is almost still reducing the visual input to a static image with a weak 
auditory noise emanating from the air conditioning vent. (B) The most salient 
location according to $\mathcal{AVSM}_{1}$ coincides with the intensity based 
salient location at the bottom of the image. (C) $\mathcal{AVSM}_{2}$ shows some 
locations that are not salient according to any feature. This may be happening 
due to exaggeration of auditory reflections which are detected as salient in 
$\mathcal{AVSM}_{2}$. (D) $\mathcal{AVSM}_{3}$ shows similar results as 
$\mathcal{AVSM}_{2}$
}
	\label{fig:AVSM-Illus-2}
\end{figure}

Consider, for example, frame \# 173 of Dataset 3, where the visual scene is 
equivalent to a static image with a weak auditory stimulus, which is the air 
conditioning vent noise (Figure~\ref{fig:AVSM-Illus-2}). Here, according to 
$\mathcal{AVSM}_{1}$ (Figure~\ref{fig:AVSM-Illus-2} (B)), the most salient 
location coincides with the strongest intensity based salient location at the 
bottom part of the image. This is because in $\mathcal{AVSM}_{1}$ we averaged 
the conspicuity maps with equal weights. So, when salient audio or motion is not 
present, the AVSM automatically switches to being a static saliency map with 
Color, Intensity and Orientation as dominant features. But, in 
$\mathcal{AVSM}_{2}$ it is computed as the average of visual and auditory 
saliency maps, hence it leads to redistribution of weights in such a way that 
each of the visual features contributes only one-eighth to the final saliency 
map and audio channel contributes one half. As a result, auditory reflections 
could get accentuated and show up as salient, which may not match with our 
judgment, as seen in Figure~\ref{fig:AVSM-Illus-2}(C).  In reality, such 
auditory reflections are imperceptible, hence may not draw our attention as 
indicated by $\mathcal{AVSM}_{2}$.

In $\mathcal{AVSM}_{3}$ where a multiplicative term, $\mathcal{VSM}(x,y,t) 
\times \mathcal{ASM}(x,y,t)$ is added, accentuates the conjunction of visual and 
auditory salient events if they are spatio-temporally coincident. But, since 
auditory and visual saliencies already contribute equally instead of the five 
independent features making equal contributions, the effect of the 
multiplicative term is small, so we see that $\mathcal{AVSM}_{3}$ has similar 
behavior as $\mathcal{AVSM}_{2}$. We did not investigate whether the conjunction 
of individual feature conspicuity maps, like $\mathcal{I}(x,y,t) \times 
\mathcal{ASM}(x,y,t)$, $\mathcal{O}(x,y,t) \times \mathcal{C}(x,y,t)$, \etc can 
result in a better saliency map. But based on visual comparison we can conclude 
that $\mathcal{AVSM}_{1}$, where each feature channel contributes equally, 
irrespective of whether it is visual or auditory, is a better AVSM computation 
method compared to $\mathcal{AVSM}_{2}$ and $\mathcal{AVSM}_{3}$. 

So an important observation we can make at this point is that, even though 
vision and audition are two separate sensory modalities and we expect them to 
equally influence the bottom-up, stimulus driven attention, this may not be the 
case. Instead, we can conclude that each feature irrespective of the sensory 
modality makes the same contribution to the final saliency map from a bottom-up 
perspective.

Next, we will compare how $\mathcal{AVSM}_{1}$ performs in comparison to 
unisensory saliency maps, namely $\mathcal{VSM}$
 and $\mathcal{ASM}$. Since $\mathcal{AVSM}_{1}$ was found to be better, the 
other two AVSMs are not discussed. 
 
\begin{figure}
	\centering
	\includegraphics[width=\linewidth]{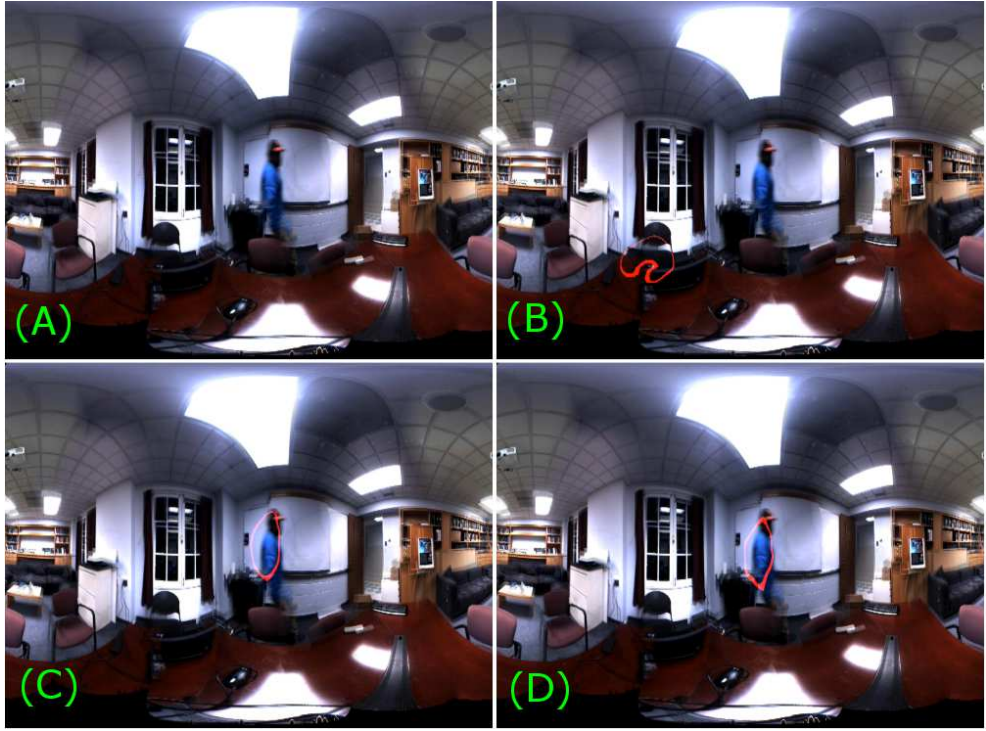}
	\caption{Comparison of AVSM with unisensory saliency maps (A) Input frame \# 50 of 
Dataset 4. The most prominent event in the scene is the person moving. (B) The 
most salient location according to $\mathcal{ASM}$ misses the most salient 
location, but shows a different location as salient (C) $\mathcal{VSM}$ shows 
the prominent motion event as salient as expected (D) $\mathcal{AVSM}_{1}$ also 
captures the salient motion event as the most salient
}
	\label{fig:AVSM-ASM-VSM-Motion}
\end{figure}

Figure~\ref{fig:AVSM-ASM-VSM-Motion} shows $\mathcal{ASM}$, $\mathcal{VSM}$ and 
$\mathcal{AVSM}_{1}$ for frame \# 50 of Dataset 4, where the person moving is 
the most salient event, which is correctly detected in $\mathcal{VSM}$ and 
$\mathcal{AVSM}_{1}$, but not in $\mathcal{ASM}$. This is expected. 

\begin{figure}
	\centering
	\includegraphics[width=\linewidth]{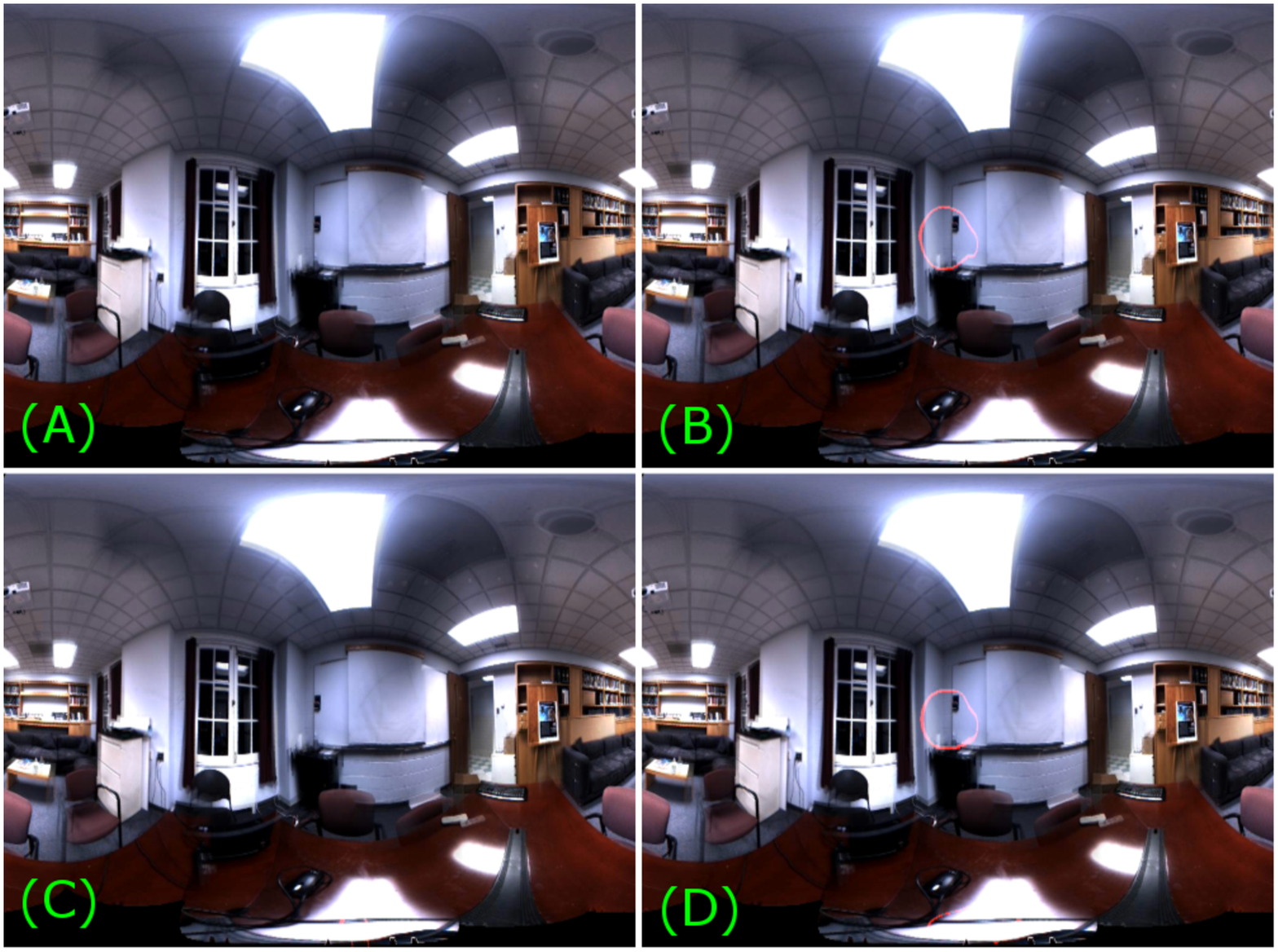}
	\caption{Comparison of AVSM with unisensory saliency maps (A) Input frame \# 346 of 
Dataset 2. The most prominent event in the scene is the audio from the 
loudspeaker. (B) The audio event is salient in $\mathcal{ASM}$ (C) 
$\mathcal{VSM}$ in this case would be equivalent to a static saliency map, hence 
the auditory salient event is missed here(D) $\mathcal{AVSM}_{1}$ also captures 
the audio as the most salient as expected
}
	\label{fig:AVSM-ASM-VSM-Audio}
\end{figure}

Next, in Figure~\ref{fig:AVSM-ASM-VSM-Audio} saliency maps for frame \# 346 of 
Dataset 2 are shown, where audio from the loudspeaker is the most salient event, 
which is correctly detected as salient in $\mathcal{ASM}$ and 
$\mathcal{AVSM}_{1}$, but missed in $\mathcal{VSM}$, which agrees with our 
judgment.

\begin{figure}
	\centering
	\includegraphics[width=\linewidth]{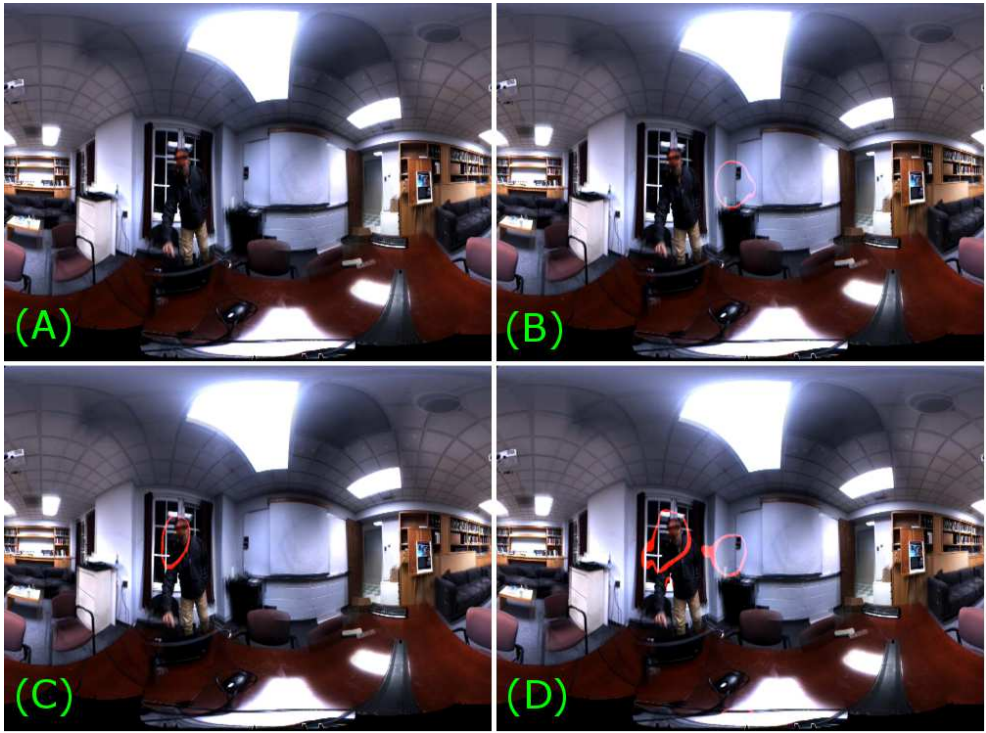}
	\caption{Comparison of AVSM with unisensory saliency maps (A) Input frame \# 393 of 
Dataset 1. (B) $\mathcal{ASM}$ detects the salient auditory event, but misses 
the salient visual motion (C) $\mathcal{VSM}$ correctly detects the salient 
motion event, but misses the salient audio. (D) $\mathcal{AVSM}_{1}$ captures 
both valid salient events from two different sensory domains
}
	\label{fig:AVSM-ASM-VS-2Events}
\end{figure}

In frame \# 393 of Dataset 1, there is strong motion of the person as well as 
sound from the loudspeaker. The unisensory and audiovisual saliency maps are 
shown in Figure~\ref{fig:AVSM-ASM-VS-2Events}. Again, the salients events 
detected by the respective saliency maps agree with our judgment. 

From these results we can conclude, the unisensory saliency maps detect valid 
unisensory events which agree with human judgment. At the same time, the 
audiovisual saliency map detects salient events from both sensory modalities, 
which again agree with our judgment. So, we can say, the AVSM detects more 
number of valid salient events compared to unisensory saliency maps. The 
unisensory saliency maps miss the salient events from the other sensory 
modality. Hence, AVSMs in general, and $\mathcal{AVSM}_{1}$ in particular, 
perform better than unisensory saliency maps in detecting valid salient events. 
As a result, AVSMs can be more useful in a variety of applications like 
surveillance, robotic navigation, \etc. Overall, the proto-object based 
audiovisual saliency map reliably detects valid salient events for all 
combinations of auditory, visual and/or audiovisual events in a majority of the 
frames. The readers can verify themselves additionally by watching the videos or 
looking at individual video frames at: 
\url{https://preview.tinyurl.com/ybg4fch4}. 

An important distinguishing factor of our AVSM computation comes from the use of 
proto-objects. With proto-object based computation, we see that salient 
locations roughly coincide with object centers giving an estimate of audiovisual 
``objectness''~\cite{alexe2012measuring}. So, this enables selection of image 
regions with possible objects based on saliency values. Moreover, since saliency 
gives a natural mechanism for ranking scene locations based on salience value, 
combined with ``objectness" that comes from proto-objects, this can serve to 
select image regions for object recognition, activity recognition, \etc with 
other methods, such as deep Convolutional Neural 
Networks~\cite{krizhevsky2012imagenet,ren2015faster}. 

Second, due to linear combination of feature conspicuity maps, the model adapts 
itself to any scene type, static or dynamic scenes, with or without audio. 
Because of this, we get a robust estimate of bottom-up saliency in a majority of 
cases. Plus, the method works well for a variety of environments, indoor and 
outdoor. 

Finally, the AVSM computed in this manner enables us to represent and compare 
saliencies of events from two different sensory modalities on a common scale. 
Other sensory modalities or feature channels can be similarly incorporated into 
the model. 

One of the factors that we have not considered in our model is the temporal 
modulation of audiovisual saliency. We treat each 100 ms interval as a snapshot, 
independent of previous frames and compute unisensory and audiovisual saliencies 
for each 100 ms frame. Even though this ``memoryless'' computation detects valid 
salient events well, temporal aspects are found to strongly influence saliency, 
especially from the auditory domain~\cite{kaya2012temporal}. Hence, factoring in 
the temporal dependence of saliency can further improve the model. For example, 
in the few cases where saliency maps appears to be noisy, we can improve the 
results with temporal smoothing of the saliency maps. Though, the proportion of 
such noisy frames is very small compared to valid detections. 

Temporal dependence of attention is important from the perspective of perception 
as well. For example, a continuous motion or an auditory alarm can be salient at 
the beginning of the event due to abrupt onset, but if it continues to persist, 
we may switch our attention to some other event, even though it is prominent in 
the scene. The mechanism and time course of multisensory attentional modulation 
needs to be further investigated and incorporated into the model. 

Another aspect, related to temporal modulation of audiovisual saliency that we 
have not considered is the Inhibition of Return (IOR)~\cite{Posner_Cohen84, 
Itti_etal98a}. IOR refers to increased reaction time to attend to a previously 
cued spatial location compared to an uncued location. The exact nature of IOR in 
the case of audiovisual attention is an active topic of 
research~\cite{van2017visually,spence1998auditory}. More recent experimental 
evidence~\cite{van2017visually} suggests that IOR is not observed in audiovisual 
attention conditions. If this is the case, not having audiovisual IOR may not be 
a significantly limiting factor, but certainly worth investigating.

Lastly, a drawback of our work is that the results are not validated with human 
psychophysics experiments. Since, saliency models aim to predict human attention 
based on bottom-up features, validating the results with human psychophysics 
experiments is necessary. Such validation would strengthen the findings of our 
study even more. But from visual judgment of the results, the readers can verify 
that the model is capable of selecting valid, perceptually salient audiovisual 
events for further processing. Moreover, our goal is to build a useful 
computational tool for automated scene analysis and the results show that the 
model is capable of doing so.
features, validating the results by presenting it human beings and measuring how 
they allocate attention in such environments, how much of their attentional 
behavior can be explained purely based on bottom-up features should be studied 
to validate and further strengthen the findings of our study. 

\section{Conclusion and Future Work}
\label{sec:proto-avsm-conclusion}
We have shown that a proto-object based audiovisual saliency map detects salient 
unisensory and multisensory events, which agree with human judgment. The AVSM 
detects a higher number of valid salient events compared to unisensory saliency 
maps demonstrating the superiority and usefulness of proto-object based 
multisensory saliency map. Among the different audiovisual saliency methods, we 
show that linear combination of individual feature channels with equal weights 
gives the best results. The AVSM computed this way performs better compared to 
others in detecting valid salient events for static as well as dynamic scenes, 
with or without salient auditory events in the scene. Also, it is less noisy and 
more robust compared to other combination methods where visual and auditory 
conspicuity maps, instead of individual feature channels, are equally weighed.

In future, incorporating the temporal modulation of saliency would be 
considered. We would also like to validate the AVSM with psychophysics 
experiments. Also, the role of Inhibition of Return in the case of audiovisual 
saliency map would also be investigated. In conclusion, a proto-object based 
audiovisual saliency map with linear and equally weighted feature channels 
detects a higher number of valid unisensory and multisensory events that agree 
with human judgment. 

\newpage

\bibliographystyle{model1-num-names}
\bibliography{thesis,refs_old,ciss2011,ciss2013avsal}

\begin{thebibliography}{76}
\expandafter\ifx\csname natexlab\endcsname\relax\def\natexlab#1{#1}\fi
\providecommand{\bibinfo}[2]{#2}
\ifx\xfnm\relax \def\xfnm[#1]{\unskip,\space#1}\fi
\bibitem[{Stein and Stanford(2008)}]{stein2008multisensory}
\bibinfo{author}{B.~E. Stein}, \bibinfo{author}{T.~R. Stanford},
\newblock \bibinfo{title}{Multisensory integration: current issues from the
  perspective of the single neuron},
\newblock \bibinfo{journal}{Nature Reviews Neuroscience} \bibinfo{volume}{9}
  (\bibinfo{year}{2008}) \bibinfo{pages}{255--266}.
\bibitem[{Stevenson and James(2009)}]{stevenson2009audiovisual}
\bibinfo{author}{R.~A. Stevenson}, \bibinfo{author}{T.~W. James},
\newblock \bibinfo{title}{Audiovisual integration in human superior temporal
  sulcus: inverse effectiveness and the neural processing of speech and object
  recognition},
\newblock \bibinfo{journal}{Neuroimage} \bibinfo{volume}{44}
  (\bibinfo{year}{2009}) \bibinfo{pages}{1210--1223}.
\bibitem[{Calvert et~al.(2004)Calvert, Spence, and Stein}]{calvert2004handbook}
\bibinfo{author}{G.~A. Calvert}, \bibinfo{author}{C.~Spence},
  \bibinfo{author}{B.~E. Stein}, \bibinfo{title}{The handbook of multisensory
  processes}, \bibinfo{publisher}{MIT press}, \bibinfo{year}{2004}.
\bibitem[{Spence et~al.(2009)Spence, Senkowski, and
  R{\"o}der}]{spence2009crossmodal}
\bibinfo{author}{C.~Spence}, \bibinfo{author}{D.~Senkowski},
  \bibinfo{author}{B.~R{\"o}der},
\newblock \bibinfo{title}{Crossmodal processing},
\newblock \bibinfo{journal}{Experimental Brain Research} \bibinfo{volume}{198}
  (\bibinfo{year}{2009}) \bibinfo{pages}{107--111}.
\bibitem[{Alais and Burr(2004)}]{alais2004ventriloquist}
\bibinfo{author}{D.~Alais}, \bibinfo{author}{D.~Burr},
\newblock \bibinfo{title}{The ventriloquist effect results from near-optimal
  bimodal integration},
\newblock \bibinfo{journal}{Current biology} \bibinfo{volume}{14}
  (\bibinfo{year}{2004}) \bibinfo{pages}{257--262}.
\bibitem[{Ghazanfar and Schroeder(2006)}]{ghazanfar2006neocortex}
\bibinfo{author}{A.~A. Ghazanfar}, \bibinfo{author}{C.~E. Schroeder},
\newblock \bibinfo{title}{Is neocortex essentially multisensory?},
\newblock \bibinfo{journal}{Trends in cognitive sciences} \bibinfo{volume}{10}
  (\bibinfo{year}{2006}) \bibinfo{pages}{278--285}.
\bibitem[{Von~Kriegstein et~al.(2005)Von~Kriegstein, Kleinschmidt, Sterzer, and
  Giraud}]{von2005interaction}
\bibinfo{author}{K.~Von~Kriegstein}, \bibinfo{author}{A.~Kleinschmidt},
  \bibinfo{author}{P.~Sterzer}, \bibinfo{author}{A.-L. Giraud},
\newblock \bibinfo{title}{Interaction of face and voice areas during speaker
  recognition},
\newblock \bibinfo{journal}{Journal of Cognitive Neuroscience}
  \bibinfo{volume}{17} (\bibinfo{year}{2005}) \bibinfo{pages}{367--376}.
\bibitem[{Watkins et~al.(2006)Watkins, Shams, Tanaka, Haynes, and
  Rees}]{watkins2006sound}
\bibinfo{author}{S.~Watkins}, \bibinfo{author}{L.~Shams},
  \bibinfo{author}{S.~Tanaka}, \bibinfo{author}{J.-D. Haynes},
  \bibinfo{author}{G.~Rees},
\newblock \bibinfo{title}{Sound alters activity in human {V1} in association
  with illusory visual perception},
\newblock \bibinfo{journal}{Neuroimage} \bibinfo{volume}{31}
  (\bibinfo{year}{2006}) \bibinfo{pages}{1247--1256}.
\bibitem[{van Wassenhove et~al.(2005)van Wassenhove, Grant, and
  Poeppel}]{van2005visual}
\bibinfo{author}{V.~van Wassenhove}, \bibinfo{author}{K.~W. Grant},
  \bibinfo{author}{D.~Poeppel},
\newblock \bibinfo{title}{Visual speech speeds up the neural processing of
  auditory speech},
\newblock \bibinfo{journal}{Proceedings of the National Academy of Sciences of
  the United States of America} \bibinfo{volume}{102} (\bibinfo{year}{2005})
  \bibinfo{pages}{1181--1186}.
\bibitem[{Ghazanfar et~al.(2005)Ghazanfar, Maier, Hoffman, and
  Logothetis}]{ghazanfar2005multisensory}
\bibinfo{author}{A.~A. Ghazanfar}, \bibinfo{author}{J.~X. Maier},
  \bibinfo{author}{K.~L. Hoffman}, \bibinfo{author}{N.~K. Logothetis},
\newblock \bibinfo{title}{Multisensory integration of dynamic faces and voices
  in rhesus monkey auditory cortex},
\newblock \bibinfo{journal}{The Journal of Neuroscience} \bibinfo{volume}{25}
  (\bibinfo{year}{2005}) \bibinfo{pages}{5004--5012}.
\bibitem[{Wang et~al.(2008)Wang, Celebrini, Trotter, and
  Barone}]{wang2008visuo}
\bibinfo{author}{Y.~Wang}, \bibinfo{author}{S.~Celebrini},
  \bibinfo{author}{Y.~Trotter}, \bibinfo{author}{P.~Barone},
\newblock \bibinfo{title}{Visuo-auditory interactions in the primary visual
  cortex of the behaving monkey: electrophysiological evidence},
\newblock \bibinfo{journal}{BMC neuroscience} \bibinfo{volume}{9}
  (\bibinfo{year}{2008}) \bibinfo{pages}{79}.
\bibitem[{{\c{C}}eting{\"u}l et~al.(2006){\c{C}}eting{\"u}l, Erzin, Yemez, and
  Tekalp}]{ccetingul2006multimodal}
\bibinfo{author}{H.~{\c{C}}eting{\"u}l}, \bibinfo{author}{E.~Erzin},
  \bibinfo{author}{Y.~Yemez}, \bibinfo{author}{A.~M. Tekalp},
\newblock \bibinfo{title}{Multimodal speaker/speech recognition using lip
  motion, lip texture and audio},
\newblock \bibinfo{journal}{Signal processing} \bibinfo{volume}{86}
  (\bibinfo{year}{2006}) \bibinfo{pages}{3549--3558}.
\bibitem[{Tamura et~al.(2005)Tamura, Iwano, and Furui}]{tamura2005toward}
\bibinfo{author}{S.~Tamura}, \bibinfo{author}{K.~Iwano},
  \bibinfo{author}{S.~Furui},
\newblock \bibinfo{title}{Toward robust multimodal speech recognition},
\newblock in: \bibinfo{booktitle}{Symposium on Large Scale Knowledge Resources
  (LKR2005)}, pp. \bibinfo{pages}{163--166}.
\bibitem[{Russell et~al.(2014)Russell, Mihala{\c{s}}, von~der Heydt, Niebur,
  and Etienne-Cummings}]{russell2014model}
\bibinfo{author}{A.~F. Russell}, \bibinfo{author}{S.~Mihala{\c{s}}},
  \bibinfo{author}{R.~von~der Heydt}, \bibinfo{author}{E.~Niebur},
  \bibinfo{author}{R.~Etienne-Cummings},
\newblock \bibinfo{title}{A model of proto-object based saliency},
\newblock \bibinfo{journal}{Vision research} \bibinfo{volume}{94}
  (\bibinfo{year}{2014}) \bibinfo{pages}{1--15}.
\bibitem[{Alais et~al.(2010)Alais, Newell, and
  Mamassian}]{alais2010multisensory}
\bibinfo{author}{D.~Alais}, \bibinfo{author}{F.~N. Newell},
  \bibinfo{author}{P.~Mamassian},
\newblock \bibinfo{title}{Multisensory processing in review: from physiology to
  behaviour},
\newblock \bibinfo{journal}{Seeing and perceiving} \bibinfo{volume}{23}
  (\bibinfo{year}{2010}) \bibinfo{pages}{3--38}.
\bibitem[{Meredith and Stein(1986)}]{meredith1986visual}
\bibinfo{author}{M.~A. Meredith}, \bibinfo{author}{B.~E. Stein},
\newblock \bibinfo{title}{Visual, auditory, and somatosensory convergence on
  cells in superior colliculus results in multisensory integration},
\newblock \bibinfo{journal}{Journal of neurophysiology} \bibinfo{volume}{56}
  (\bibinfo{year}{1986}) \bibinfo{pages}{640--662}.
\bibitem[{Evangelopoulos et~al.(2008)Evangelopoulos, Rapantzikos, Potamianos,
  Maragos, Zlatintsi, and Avrithis}]{evangelopoulos2008movie}
\bibinfo{author}{G.~Evangelopoulos}, \bibinfo{author}{K.~Rapantzikos},
  \bibinfo{author}{A.~Potamianos}, \bibinfo{author}{P.~Maragos},
  \bibinfo{author}{A.~Zlatintsi}, \bibinfo{author}{Y.~Avrithis},
\newblock \bibinfo{title}{Movie summarization based on audiovisual saliency
  detection},
\newblock in: \bibinfo{booktitle}{15th IEEE International Conference on Image
  Processing}, \bibinfo{organization}{IEEE}, pp. \bibinfo{pages}{2528--2531}.
\bibitem[{Song(2013)}]{song2013effet}
\bibinfo{author}{G.~Song}, \bibinfo{title}{Effet du son dans les vid{\'e}os sur
  la direction du regard: contribution {\`a} la mod{\'e}lisation de la
  saillance audiovisuelle}, Ph.D. thesis, Universit{\'e} de Grenoble,
  \bibinfo{year}{2013}.
\bibitem[{Ramenahalli et~al.(2013)Ramenahalli, Mendat, Dura-Bernal,
  Culurciello, Nieburt, and Andreou}]{ramenahalli2013audio}
\bibinfo{author}{S.~Ramenahalli}, \bibinfo{author}{D.~R. Mendat},
  \bibinfo{author}{S.~Dura-Bernal}, \bibinfo{author}{E.~Culurciello},
  \bibinfo{author}{E.~Nieburt}, \bibinfo{author}{A.~Andreou},
\newblock \bibinfo{title}{Audio-visual saliency map: overview, basic models and
  hardware implementation},
\newblock in: \bibinfo{booktitle}{Information Sciences and Systems (CISS), 2013
  47th Annual Conference on}, \bibinfo{organization}{IEEE}, pp.
  \bibinfo{pages}{1--6}.
\bibitem[{Grossberg et~al.(1997)Grossberg, Roberts, Aguilar, and
  Bullock}]{grossberg1997neural}
\bibinfo{author}{S.~Grossberg}, \bibinfo{author}{K.~Roberts},
  \bibinfo{author}{M.~Aguilar}, \bibinfo{author}{D.~Bullock},
\newblock \bibinfo{title}{A neural model of multimodal adaptive saccadic eye
  movement control by superior colliculus},
\newblock \bibinfo{journal}{The Journal of neuroscience} \bibinfo{volume}{17}
  (\bibinfo{year}{1997}) \bibinfo{pages}{9706--9725}.
\bibitem[{Meredith and Stein(1996)}]{meredith1996spatial}
\bibinfo{author}{M.~A. Meredith}, \bibinfo{author}{B.~E. Stein},
\newblock \bibinfo{title}{Spatial determinants of multisensory integration in
  cat superior colliculus neurons},
\newblock \bibinfo{journal}{Journal of Neurophysiology} \bibinfo{volume}{75}
  (\bibinfo{year}{1996}) \bibinfo{pages}{1843--1857}.
\bibitem[{Meredith et~al.(1987)Meredith, Nemitz, and
  Stein}]{meredith1987determinants}
\bibinfo{author}{M.~A. Meredith}, \bibinfo{author}{J.~W. Nemitz},
  \bibinfo{author}{B.~E. Stein},
\newblock \bibinfo{title}{Determinants of multisensory integration in superior
  colliculus neurons. {I}. temporal factors},
\newblock \bibinfo{journal}{The Journal of neuroscience} \bibinfo{volume}{7}
  (\bibinfo{year}{1987}) \bibinfo{pages}{3215--3229}.
\bibitem[{Casey et~al.(2012)Casey, Pavlou, and Timotheou}]{casey2012audio}
\bibinfo{author}{M.~C. Casey}, \bibinfo{author}{A.~Pavlou},
  \bibinfo{author}{A.~Timotheou},
\newblock \bibinfo{title}{Audio-visual localization with hierarchical
  topographic maps: Modeling the superior colliculus},
\newblock \bibinfo{journal}{Neurocomputing} \bibinfo{volume}{97}
  (\bibinfo{year}{2012}) \bibinfo{pages}{344--356}.
\bibitem[{Huo and Murray(2009)}]{huo2009adaptation}
\bibinfo{author}{J.~Huo}, \bibinfo{author}{A.~Murray},
\newblock \bibinfo{title}{The adaptation of visual and auditory integration in
  the barn owl superior colliculus with spike timing dependent plasticity},
\newblock \bibinfo{journal}{Neural Networks} \bibinfo{volume}{22}
  (\bibinfo{year}{2009}) \bibinfo{pages}{913--921}.
\bibitem[{Huo et~al.(2012)Huo, Murray, and Wei}]{huo2012adaptive}
\bibinfo{author}{J.~Huo}, \bibinfo{author}{A.~Murray},
  \bibinfo{author}{D.~Wei},
\newblock \bibinfo{title}{Adaptive visual and auditory map alignment in barn
  owl superior colliculus and its neuromorphic implementation},
\newblock \bibinfo{journal}{Neural Networks and Learning Systems, IEEE
  Transactions on} \bibinfo{volume}{23} (\bibinfo{year}{2012})
  \bibinfo{pages}{1486--1497}.
\bibitem[{Anastasio et~al.(2000)Anastasio, Patton, and
  Belkacem-Boussaid}]{anastasio2000using}
\bibinfo{author}{T.~J. Anastasio}, \bibinfo{author}{P.~E. Patton},
  \bibinfo{author}{K.~Belkacem-Boussaid},
\newblock \bibinfo{title}{Using bayes' rule to model multisensory enhancement
  in the superior colliculus},
\newblock \bibinfo{journal}{Neural Computation} \bibinfo{volume}{12}
  (\bibinfo{year}{2000}) \bibinfo{pages}{1165--1187}.
\bibitem[{Patton et~al.(2002)Patton, Belkacem-Boussaid, and
  Anastasio}]{patton2002multimodality}
\bibinfo{author}{P.~Patton}, \bibinfo{author}{K.~Belkacem-Boussaid},
  \bibinfo{author}{T.~J. Anastasio},
\newblock \bibinfo{title}{Multimodality in the superior colliculus: an
  information theoretic analysis},
\newblock \bibinfo{journal}{Cognitive Brain Research} \bibinfo{volume}{14}
  (\bibinfo{year}{2002}) \bibinfo{pages}{10--19}.
\bibitem[{Patton and Anastasio(2003)}]{patton2003modeling}
\bibinfo{author}{P.~E. Patton}, \bibinfo{author}{T.~J. Anastasio},
\newblock \bibinfo{title}{Modeling cross-modal enhancement and
  modality-specific suppression in multisensory neurons},
\newblock \bibinfo{journal}{Neural computation} \bibinfo{volume}{15}
  (\bibinfo{year}{2003}) \bibinfo{pages}{783--810}.
\bibitem[{Colonius and Diederich(2004)}]{colonius2004aren}
\bibinfo{author}{H.~Colonius}, \bibinfo{author}{A.~Diederich},
\newblock \bibinfo{title}{Why aren’t all deep superior colliculus neurons
  multisensory? a bayes’ ratio analysis},
\newblock \bibinfo{journal}{Cognitive, Affective, \& Behavioral Neuroscience}
  \bibinfo{volume}{4} (\bibinfo{year}{2004}) \bibinfo{pages}{344--353}.
\bibitem[{Ma et~al.(2006)Ma, Beck, Latham, and Pouget}]{ma2006bayesian}
\bibinfo{author}{W.~J. Ma}, \bibinfo{author}{J.~M. Beck},
  \bibinfo{author}{P.~E. Latham}, \bibinfo{author}{A.~Pouget},
\newblock \bibinfo{title}{Bayesian inference with probabilistic population
  codes},
\newblock \bibinfo{journal}{Nature neuroscience} \bibinfo{volume}{9}
  (\bibinfo{year}{2006}) \bibinfo{pages}{1432--1438}.
\bibitem[{Wilson et~al.(2002)Wilson, Rangarajan, Checka, and
  Darrell}]{Wilson2002}
\bibinfo{author}{K.~Wilson}, \bibinfo{author}{V.~Rangarajan},
  \bibinfo{author}{N.~Checka}, \bibinfo{author}{T.~Darrell},
\newblock \bibinfo{title}{Audiovisual arrays for untethered spoken interfaces},
\newblock in: \bibinfo{booktitle}{Proceedings of the 4th IEEE International
  Conference on Multimodal Interfaces}, ICMI '02, \bibinfo{publisher}{IEEE
  Computer Society}, \bibinfo{address}{Washington, DC, USA},
  \bibinfo{year}{2002}, pp. \bibinfo{pages}{389--}.
\bibitem[{Torres and Kalva(2014)}]{torres2014influence}
\bibinfo{author}{F.~Torres}, \bibinfo{author}{H.~Kalva},
\newblock \bibinfo{title}{Influence of audio triggered emotional attention on
  video perception},
\newblock in: \bibinfo{booktitle}{IS\&T/SPIE Electronic Imaging},
  \bibinfo{organization}{International Society for Optics and Photonics}, pp.
  \bibinfo{pages}{901408--901408}.
\bibitem[{Lee et~al.(2011)Lee, De~Simone, and Ebrahimi}]{lee2011efficient}
\bibinfo{author}{J.-S. Lee}, \bibinfo{author}{F.~De~Simone},
  \bibinfo{author}{T.~Ebrahimi},
\newblock \bibinfo{title}{Efficient video coding based on audio-visual focus of
  attention},
\newblock \bibinfo{journal}{Journal of Visual Communication and Image
  Representation} \bibinfo{volume}{22} (\bibinfo{year}{2011})
  \bibinfo{pages}{704--711}.
\bibitem[{Rerabek et~al.(2014)Rerabek, Nemoto, Lee, and
  Ebrahimi}]{rerabek2014audiovisual}
\bibinfo{author}{M.~Rerabek}, \bibinfo{author}{H.~Nemoto},
  \bibinfo{author}{J.-S. Lee}, \bibinfo{author}{T.~Ebrahimi},
\newblock \bibinfo{title}{Audiovisual focus of attention and its application to
  ultra high definition video compression},
\newblock in: \bibinfo{booktitle}{IS\&T/SPIE Electronic Imaging},
  \bibinfo{organization}{International Society for Optics and Photonics}, pp.
  \bibinfo{pages}{901407--901407}.
\bibitem[{Ruesch et~al.(2008)Ruesch, Lopes, Bernardino, Hornstein,
  Santos-Victor, and Pfeifer}]{ruesch2008multimodal}
\bibinfo{author}{J.~Ruesch}, \bibinfo{author}{M.~Lopes},
  \bibinfo{author}{A.~Bernardino}, \bibinfo{author}{J.~Hornstein},
  \bibinfo{author}{J.~Santos-Victor}, \bibinfo{author}{R.~Pfeifer},
\newblock \bibinfo{title}{Multimodal saliency-based bottom-up attention a
  framework for the humanoid robot icub},
\newblock in: \bibinfo{booktitle}{IEEE International Conference on Robotics and
  Automation}, \bibinfo{organization}{IEEE}, pp. \bibinfo{pages}{962--967}.
\bibitem[{Schauerte et~al.(2009)Schauerte, Richarz, Pl{\"o}tz, Thurau, and
  Fink}]{schauerte2009multi}
\bibinfo{author}{B.~Schauerte}, \bibinfo{author}{J.~Richarz},
  \bibinfo{author}{T.~Pl{\"o}tz}, \bibinfo{author}{C.~Thurau},
  \bibinfo{author}{G.~A. Fink},
\newblock \bibinfo{title}{Multi-modal and multi-camera attention in smart
  environments},
\newblock in: \bibinfo{booktitle}{Proceedings of the 2009 international
  conference on Multimodal interfaces}, \bibinfo{organization}{ACM}, pp.
  \bibinfo{pages}{261--268}.
\bibitem[{Schauerte et~al.(2011)Schauerte, Kuhn, Kroschel, and
  Stiefelhagen}]{schauerte2011multimodal}
\bibinfo{author}{B.~Schauerte}, \bibinfo{author}{B.~Kuhn},
  \bibinfo{author}{K.~Kroschel}, \bibinfo{author}{R.~Stiefelhagen},
\newblock \bibinfo{title}{Multimodal saliency-based attention for object-based
  scene analysis},
\newblock in: \bibinfo{booktitle}{IEEE/RSJ International Conference on
  Intelligent Robots and Systems}, \bibinfo{organization}{IEEE}, pp.
  \bibinfo{pages}{1173--1179}.
\bibitem[{Schauerte(2016)}]{schauerte2016bottom}
\bibinfo{author}{B.~Schauerte},
\newblock \bibinfo{title}{Bottom-up audio-visual attention for scene
  exploration},
\newblock in: \bibinfo{booktitle}{Multimodal Computational Attention for Scene
  Understanding and Robotics}, \bibinfo{publisher}{Springer},
  \bibinfo{year}{2016}, pp. \bibinfo{pages}{35--113}.
\bibitem[{Onat et~al.(2007)Onat, Libertus, and Konig}]{2007OnatIntAVovertAtt}
\bibinfo{author}{S.~Onat}, \bibinfo{author}{K.~Libertus},
  \bibinfo{author}{P.~Konig},
\newblock \bibinfo{title}{Integrating audiovisual information for the control
  of overt attention},
\newblock in: \bibinfo{booktitle}{Journal of Vision, 7(10):11},
  \bibinfo{year}{2007}.
\bibitem[{K{\"u}hn et~al.(2012{\natexlab{a}})K{\"u}hn, Schauerte, Stiefelhagen,
  and Kroschel}]{kuhn2012modular}
\bibinfo{author}{B.~K{\"u}hn}, \bibinfo{author}{B.~Schauerte},
  \bibinfo{author}{R.~Stiefelhagen}, \bibinfo{author}{K.~Kroschel},
\newblock \bibinfo{title}{A modular audio-visual scene analysis and attention
  system for humanoid robots},
\newblock in: \bibinfo{booktitle}{Proc. 43rd Int. Symp. Robotics (ISR)}.
\bibitem[{K{\"u}hn et~al.(2012{\natexlab{b}})K{\"u}hn, Schauerte, Kroschel, and
  Stiefelhagen}]{kuhn2012multimodal}
\bibinfo{author}{B.~K{\"u}hn}, \bibinfo{author}{B.~Schauerte},
  \bibinfo{author}{K.~Kroschel}, \bibinfo{author}{R.~Stiefelhagen},
\newblock \bibinfo{title}{Multimodal saliency-based attention: A lazy robot's
  approach},
\newblock in: \bibinfo{booktitle}{IEEE/RSJ International Conference on
  Intelligent Robots and Systems}, \bibinfo{organization}{IEEE}, pp.
  \bibinfo{pages}{807--814}.
\bibitem[{Bauer et~al.(2012)Bauer, Weber, and Wermter}]{bauer2012som}
\bibinfo{author}{J.~Bauer}, \bibinfo{author}{C.~Weber},
  \bibinfo{author}{S.~Wermter},
\newblock \bibinfo{title}{A som-based model for multi-sensory integration in
  the superior colliculus},
\newblock in: \bibinfo{booktitle}{The 2012 international joint conference on
  Neural Networks (IJCNN)}, \bibinfo{organization}{IEEE}, pp.
  \bibinfo{pages}{1--8}.
\bibitem[{Viciana-Abad et~al.(2014)Viciana-Abad, Marfil, Perez-Lorenzo,
  Bandera, Romero-Garces, and Reche-Lopez}]{viciana2014audio}
\bibinfo{author}{R.~Viciana-Abad}, \bibinfo{author}{R.~Marfil},
  \bibinfo{author}{J.~M. Perez-Lorenzo}, \bibinfo{author}{J.~P. Bandera},
  \bibinfo{author}{A.~Romero-Garces}, \bibinfo{author}{P.~Reche-Lopez},
\newblock \bibinfo{title}{Audio-visual perception system for a humanoid robotic
  head},
\newblock \bibinfo{journal}{Sensors} \bibinfo{volume}{14}
  (\bibinfo{year}{2014}) \bibinfo{pages}{9522--9545}.
\bibitem[{Evangelopoulos et~al.(2008)Evangelopoulos, Rapantzikos, Maragos,
  Avrithis, and Potamianos}]{evangelopoulos2008audiovisual}
\bibinfo{author}{G.~Evangelopoulos}, \bibinfo{author}{K.~Rapantzikos},
  \bibinfo{author}{P.~Maragos}, \bibinfo{author}{Y.~Avrithis},
  \bibinfo{author}{A.~Potamianos},
\newblock \bibinfo{title}{Audiovisual attention modeling and salient event
  detection},
\newblock in: \bibinfo{booktitle}{Multimodal Processing and Interaction},
  \bibinfo{publisher}{Springer}, \bibinfo{year}{2008}, pp.
  \bibinfo{pages}{1--21}.
\bibitem[{Rapantzikos et~al.(2007)Rapantzikos, Evangelopoulos, Maragos, and
  Avrithis}]{rapantzikos2007audio}
\bibinfo{author}{K.~Rapantzikos}, \bibinfo{author}{G.~Evangelopoulos},
  \bibinfo{author}{P.~Maragos}, \bibinfo{author}{Y.~Avrithis},
\newblock \bibinfo{title}{An audio-visual saliency model for movie
  summarization},
\newblock in: \bibinfo{booktitle}{Multimedia Signal Processing, 2007. MMSP
  2007. IEEE 9th Workshop on}, \bibinfo{organization}{IEEE}, pp.
  \bibinfo{pages}{320--323}.
\bibitem[{Evangelopoulos et~al.(2013)Evangelopoulos, Zlatintsi, Potamianos,
  Maragos, Rapantzikos, Skoumas, and Avrithis}]{evangelopoulos2013multimodal}
\bibinfo{author}{G.~Evangelopoulos}, \bibinfo{author}{A.~Zlatintsi},
  \bibinfo{author}{A.~Potamianos}, \bibinfo{author}{P.~Maragos},
  \bibinfo{author}{K.~Rapantzikos}, \bibinfo{author}{G.~Skoumas},
  \bibinfo{author}{Y.~Avrithis},
\newblock \bibinfo{title}{Multimodal saliency and fusion for movie
  summarization based on aural, visual, and textual attention},
\newblock \bibinfo{journal}{Multimedia, IEEE Transactions on}
  \bibinfo{volume}{15} (\bibinfo{year}{2013}) \bibinfo{pages}{1553--1568}.
\bibitem[{Nakajima et~al.(2014)Nakajima, Sugimoto, and
  Kawamoto}]{nakajima2014incorporating}
\bibinfo{author}{J.~Nakajima}, \bibinfo{author}{A.~Sugimoto},
  \bibinfo{author}{K.~Kawamoto},
\newblock \bibinfo{title}{Incorporating audio signals into constructing a
  visual saliency map},
\newblock in: \bibinfo{booktitle}{Image and Video Technology},
  \bibinfo{publisher}{Springer}, \bibinfo{year}{2014}, pp.
  \bibinfo{pages}{468--480}.
\bibitem[{Itti and Baldi(2008)}]{Itti_Baldi08}
\bibinfo{author}{L.~Itti}, \bibinfo{author}{P.~Baldi},
\newblock \bibinfo{title}{Bayesian surprise attracts human attention},
\newblock \bibinfo{journal}{Vision Research} \bibinfo{volume}{epub}
  (\bibinfo{year}{2008}).
\bibitem[{Nakajima et~al.(2015)Nakajima, Kimura, Sugimoto, and
  Kashino}]{nakajima2015visual}
\bibinfo{author}{J.~Nakajima}, \bibinfo{author}{A.~Kimura},
  \bibinfo{author}{A.~Sugimoto}, \bibinfo{author}{K.~Kashino},
\newblock \bibinfo{title}{Visual attention driven by auditory cues},
\newblock in: \bibinfo{booktitle}{MultiMedia Modeling},
  \bibinfo{organization}{Springer}, pp. \bibinfo{pages}{74--86}.
\bibitem[{Korchagin et~al.(2011)Korchagin, Motlicek, Duffner, and
  Bourlard}]{korchagin2011just}
\bibinfo{author}{D.~Korchagin}, \bibinfo{author}{P.~Motlicek},
  \bibinfo{author}{S.~Duffner}, \bibinfo{author}{H.~Bourlard},
\newblock \bibinfo{title}{Just-in-time multimodal association and fusion from
  home entertainment},
\newblock in: \bibinfo{booktitle}{2011 IEEE International Conference on
  Multimedia and Expo (ICME)}, \bibinfo{organization}{IEEE}, pp.
  \bibinfo{pages}{1--5}.
\bibitem[{Hershey and Movellan(2000)}]{hershey2000audio}
\bibinfo{author}{J.~R. Hershey}, \bibinfo{author}{J.~R. Movellan},
\newblock \bibinfo{title}{Audio vision: Using audio-visual synchrony to locate
  sounds},
\newblock in: \bibinfo{booktitle}{Advances in Neural Information Processing
  Systems}, pp. \bibinfo{pages}{813--819}.
\bibitem[{Blauth et~al.(2012)Blauth, Minotto, Jung, Lee, and
  Kalker}]{blauth2012voice}
\bibinfo{author}{D.~A. Blauth}, \bibinfo{author}{V.~P. Minotto},
  \bibinfo{author}{C.~R. Jung}, \bibinfo{author}{B.~Lee},
  \bibinfo{author}{T.~Kalker},
\newblock \bibinfo{title}{Voice activity detection and speaker localization
  using audiovisual cues},
\newblock \bibinfo{journal}{Pattern Recognition Letters} \bibinfo{volume}{33}
  (\bibinfo{year}{2012}) \bibinfo{pages}{373--380}.
\bibitem[{Ratajczak et~al.(2016)Ratajczak, Pellerin, Labourey, and
  Garbay}]{ratajczak2016fast}
\bibinfo{author}{R.~Ratajczak}, \bibinfo{author}{D.~Pellerin},
  \bibinfo{author}{Q.~Labourey}, \bibinfo{author}{C.~Garbay},
\newblock \bibinfo{title}{A fast audiovisual attention model for human
  detection and localization on a companion robot},
\newblock in: \bibinfo{booktitle}{The First International Conference on
  Applications and Systems of Visual Paradigms (VISUAL 2016)}.
\bibitem[{Song et~al.(2012)Song, Pellerin, and Granjon}]{song2012different}
\bibinfo{author}{G.~Song}, \bibinfo{author}{D.~Pellerin},
  \bibinfo{author}{L.~Granjon},
\newblock \bibinfo{title}{How different kinds of sound in videos can influence
  gaze},
\newblock in: \bibinfo{booktitle}{13th International Workshop on Image Analysis
  for Multimedia Interactive Services (WIAMIS)}, \bibinfo{organization}{IEEE},
  pp. \bibinfo{pages}{1--4}.
\bibitem[{Coutrot and Guyader(2014{\natexlab{a}})}]{coutrot2014saliency}
\bibinfo{author}{A.~Coutrot}, \bibinfo{author}{N.~Guyader},
\newblock \bibinfo{title}{How saliency, faces, and sound influence gaze in
  dynamic social scenes},
\newblock \bibinfo{journal}{Journal of Vision} \bibinfo{volume}{14}
  (\bibinfo{year}{2014}{\natexlab{a}}) \bibinfo{pages}{5}.
\bibitem[{Coutrot and Guyader(2014{\natexlab{b}})}]{coutrot2014audiovisual}
\bibinfo{author}{A.~Coutrot}, \bibinfo{author}{N.~Guyader},
\newblock \bibinfo{title}{An audiovisual attention model for natural
  conversation scenes},
\newblock in: \bibinfo{booktitle}{IEEE International Conference on Image
  Processing (ICIP)}, \bibinfo{organization}{IEEE}, pp.
  \bibinfo{pages}{1100--1104}.
\bibitem[{Noulas et~al.(2012)Noulas, Englebienne, and
  Kr{\"o}se}]{noulas2012multimodal}
\bibinfo{author}{A.~Noulas}, \bibinfo{author}{G.~Englebienne},
  \bibinfo{author}{B.~J. Kr{\"o}se},
\newblock \bibinfo{title}{Multimodal speaker diarization},
\newblock \bibinfo{journal}{IEEE Transactions on Pattern Analysis and Machine
  Intelligence} \bibinfo{volume}{34} (\bibinfo{year}{2012})
  \bibinfo{pages}{79--93}.
\bibitem[{Miro et~al.(2012)Miro, Bozonnet, Evans, Fredouille, Friedland, and
  Vinyals}]{miro2012speaker}
\bibinfo{author}{X.~A. Miro}, \bibinfo{author}{S.~Bozonnet},
  \bibinfo{author}{N.~Evans}, \bibinfo{author}{C.~Fredouille},
  \bibinfo{author}{G.~Friedland}, \bibinfo{author}{O.~Vinyals},
\newblock \bibinfo{title}{Speaker diarization: A review of recent research},
\newblock \bibinfo{journal}{IEEE Transactions on Audio, Speech, and Language
  Processing} \bibinfo{volume}{20} (\bibinfo{year}{2012})
  \bibinfo{pages}{356--370}.
\bibitem[{Sidaty et~al.(2014)Sidaty, Larabi, and Saadane}]{sidaty2014towards}
\bibinfo{author}{N.~O. Sidaty}, \bibinfo{author}{M.-C. Larabi},
  \bibinfo{author}{A.~Saadane},
\newblock \bibinfo{title}{Towards understanding and modeling audiovisual
  saliency based on talking faces},
\newblock in: \bibinfo{booktitle}{Tenth International Conference on
  Signal-Image Technology and Internet-Based Systems (SITIS)},
  \bibinfo{organization}{IEEE}, pp. \bibinfo{pages}{508--515}.
\bibitem[{Russell et~al.(2014)Russell, Mihalas, von~der Heydt, Niebur, and
  Etienne-Cummings}]{Russell_etal14}
\bibinfo{author}{A.~F. Russell}, \bibinfo{author}{S.~Mihalas},
  \bibinfo{author}{R.~von~der Heydt}, \bibinfo{author}{E.~Niebur},
  \bibinfo{author}{R.~Etienne-Cummings},
\newblock \bibinfo{title}{A model of proto-object based saliency},
\newblock \bibinfo{journal}{Vision Research} \bibinfo{volume}{94}
  (\bibinfo{year}{2014}) \bibinfo{pages}{1--15}.
\bibitem[{Rensink(2000)}]{rensink2000dynamic}
\bibinfo{author}{R.~A. Rensink},
\newblock \bibinfo{title}{The dynamic representation of scenes},
\newblock \bibinfo{journal}{Visual cognition} \bibinfo{volume}{7}
  (\bibinfo{year}{2000}) \bibinfo{pages}{17--42}.
\bibitem[{Sun et~al.(2010)Sun, Roth, and Black}]{sun2010secrets}
\bibinfo{author}{D.~Sun}, \bibinfo{author}{S.~Roth}, \bibinfo{author}{M.~J.
  Black},
\newblock \bibinfo{title}{Secrets of optical flow estimation and their
  principles},
\newblock in: \bibinfo{booktitle}{2010 IEEE Conference on Computer Vision and
  Pattern Recognition (CVPR)}, \bibinfo{organization}{IEEE}, pp.
  \bibinfo{pages}{2432--2439}.
\bibitem[{Sun et~al.(2008)Sun, Roth, and Black}]{opticFlowDSun}
\bibinfo{author}{D.~Sun}, \bibinfo{author}{S.~Roth},
  \bibinfo{author}{M.~Black}, \bibinfo{title}{Optic flow estimation matlab
  code},
  \bibinfo{howpublished}{\url{http://cs.brown.edu/~dqsun/research/software.html}},
  \bibinfo{year}{2008}.
\bibitem[{Molin et~al.(2013)Molin, Russell, Mihalas, Niebur, and
  Etienne-Cummings}]{molin2013proto}
\bibinfo{author}{J.~L. Molin}, \bibinfo{author}{A.~F. Russell},
  \bibinfo{author}{S.~Mihalas}, \bibinfo{author}{E.~Niebur},
  \bibinfo{author}{R.~Etienne-Cummings},
\newblock \bibinfo{title}{Proto-object based visual saliency model with a
  motion-sensitive channel},
\newblock in: \bibinfo{booktitle}{Biomedical Circuits and Systems Conference
  (BioCAS), 2013 IEEE}, pp. \bibinfo{pages}{25--28}.
\bibitem[{Donovan et~al.(2007)Donovan, Duraiswami, and
  Neumann}]{donovan2007microphone}
\bibinfo{author}{A.~O. Donovan}, \bibinfo{author}{R.~Duraiswami},
  \bibinfo{author}{J.~Neumann},
\newblock \bibinfo{title}{Microphone arrays as generalized cameras for
  integrated audio visual processing},
\newblock in: \bibinfo{booktitle}{IEEE Conference on Computer Vision and
  Pattern Recognition}, \bibinfo{organization}{IEEE}, pp.
  \bibinfo{pages}{1--8}.
\bibitem[{Meyer and Elko(2002)}]{meyer2002highly}
\bibinfo{author}{J.~Meyer}, \bibinfo{author}{G.~Elko},
\newblock \bibinfo{title}{A highly scalable spherical microphone array based on
  an orthonormal decomposition of the soundfield},
\newblock in: \bibinfo{booktitle}{IEEE International Conference on Acoustics,
  Speech, and Signal Processing (ICASSP)}, volume~\bibinfo{volume}{2},
  \bibinfo{organization}{IEEE}, pp. \bibinfo{pages}{II--1781}.
\bibitem[{O'Donovan et~al.(2007)O'Donovan, Duraiswami, Gumerov
  et~al.}]{o2007real}
\bibinfo{author}{A.~O'Donovan}, \bibinfo{author}{R.~Duraiswami},
  \bibinfo{author}{N.~Gumerov}, et~al.,
\newblock \bibinfo{title}{Real time capture of audio images and their use with
  video},
\newblock in: \bibinfo{booktitle}{IEEE Workshop on Applications of Signal
  Processing to Audio and Acoustics}, \bibinfo{organization}{IEEE}, pp.
  \bibinfo{pages}{10--13}.
\bibitem[{Zhang and von~der Heydt(2010)}]{Zhang_vonderHeydt10}
\bibinfo{author}{N.~Zhang}, \bibinfo{author}{R.~von~der Heydt},
\newblock \bibinfo{title}{Analysis of the context integration mechanisms
  underlying figure--ground organization in the visual cortex},
\newblock \bibinfo{journal}{The Journal of Neuroscience} \bibinfo{volume}{30}
  (\bibinfo{year}{2010}) \bibinfo{pages}{6482--6496}.
\bibitem[{Itti et~al.(1998)Itti, Koch, and Niebur}]{Itti_etal98a}
\bibinfo{author}{L.~Itti}, \bibinfo{author}{C.~Koch},
  \bibinfo{author}{E.~Niebur},
\newblock \bibinfo{title}{A model of saliency-based fast visual attention for
  rapid scene analysis},
\newblock \bibinfo{journal}{IEEE Transactions on Pattern Analysis and Machine
  Intelligence} \bibinfo{volume}{20} (\bibinfo{year}{1998})
  \bibinfo{pages}{1254--1259}.
\bibitem[{Alexe et~al.(2012)Alexe, Deselaers, and Ferrari}]{alexe2012measuring}
\bibinfo{author}{B.~Alexe}, \bibinfo{author}{T.~Deselaers},
  \bibinfo{author}{V.~Ferrari},
\newblock \bibinfo{title}{Measuring the objectness of image windows},
\newblock \bibinfo{journal}{IEEE transactions on pattern analysis and machine
  intelligence} \bibinfo{volume}{34} (\bibinfo{year}{2012})
  \bibinfo{pages}{2189--2202}.
\bibitem[{Krizhevsky et~al.(2012)Krizhevsky, Sutskever, and
  Hinton}]{krizhevsky2012imagenet}
\bibinfo{author}{A.~Krizhevsky}, \bibinfo{author}{I.~Sutskever},
  \bibinfo{author}{G.~E. Hinton},
\newblock \bibinfo{title}{Imagenet classification with deep convolutional
  neural networks},
\newblock in: \bibinfo{booktitle}{Advances in neural information processing
  systems}, pp. \bibinfo{pages}{1097--1105}.
\bibitem[{Ren et~al.(2015)Ren, He, Girshick, and Sun}]{ren2015faster}
\bibinfo{author}{S.~Ren}, \bibinfo{author}{K.~He},
  \bibinfo{author}{R.~Girshick}, \bibinfo{author}{J.~Sun},
\newblock \bibinfo{title}{Faster {R-CNN}: Towards real-time object detection
  with region proposal networks},
\newblock in: \bibinfo{booktitle}{Advances in neural information processing
  systems}, pp. \bibinfo{pages}{91--99}.
\bibitem[{Kaya and Elhilali(2012)}]{kaya2012temporal}
\bibinfo{author}{E.~M. Kaya}, \bibinfo{author}{M.~Elhilali},
\newblock \bibinfo{title}{A temporal saliency map for modeling auditory
  attention},
\newblock in: \bibinfo{booktitle}{Information Sciences and Systems (CISS), 2012
  46th Annual Conference on}, \bibinfo{organization}{IEEE}, pp.
  \bibinfo{pages}{1--6}.
\bibitem[{Posner and Cohen(1984)}]{Posner_Cohen84}
\bibinfo{author}{M.~I. Posner}, \bibinfo{author}{Y.~Cohen},
\newblock \bibinfo{title}{Components of visual orienting},
\newblock in: \bibinfo{editor}{H.~Bouma}, \bibinfo{editor}{D.~G. Bouwhuis}
  (Eds.), \bibinfo{booktitle}{Attention and Performance {X}},
  \bibinfo{publisher}{Hilldale}, \bibinfo{address}{NJ}, \bibinfo{year}{1984},
  pp. \bibinfo{pages}{531--556}.
\bibitem[{Van~der Stoep et~al.(2017)Van~der Stoep, Van~der Stigchel, Nijboer,
  and Spence}]{van2017visually}
\bibinfo{author}{N.~Van~der Stoep}, \bibinfo{author}{S.~Van~der Stigchel},
  \bibinfo{author}{T.~Nijboer}, \bibinfo{author}{C.~Spence},
\newblock \bibinfo{title}{Visually induced inhibition of return affects the
  integration of auditory and visual information},
\newblock \bibinfo{journal}{Perception} \bibinfo{volume}{46}
  (\bibinfo{year}{2017}) \bibinfo{pages}{6--17}.
\bibitem[{Spence and Driver(1998)}]{spence1998auditory}
\bibinfo{author}{C.~Spence}, \bibinfo{author}{J.~Driver},
\newblock \bibinfo{title}{Auditory and audiovisual inhibition of return},
\newblock \bibinfo{journal}{Attention, Perception, \& Psychophysics}
  \bibinfo{volume}{60} (\bibinfo{year}{1998}) \bibinfo{pages}{125--139}.

\end{thebibliography}
\newpage
\end{document}